\documentclass[9pt,showpacs,preprintnumbers,superscriptaddress,amsmath,amssymb,nofootinbib]{revtex4-2}
\usepackage{amsmath, graphicx}
\usepackage{dcolumn}
\usepackage{bm}
\usepackage{color}
\usepackage{epstopdf}
\usepackage{eurosym}
\usepackage{multirow}
\usepackage{amsfonts}
\usepackage{amsmath}
\usepackage{hyperref}
\usepackage{amssymb}
\usepackage[english]{babel}
\usepackage{graphicx}
\usepackage{epsfig}
\usepackage{subfigure}
\usepackage{unicode}
\begin{document}
\newcommand{\hs}{\hspace*{0.2cm}}
\newcommand{\hsp}{\hspace*{0.5cm}}
\newcommand{\vs}{\vspace*{0.5cm}}
\newcommand{\be}{\begin{equation}}
\newcommand{\ee}{\end{equation}}
\newcommand{\bea}{\begin{eqnarray}}
\newcommand{\eea}{\end{eqnarray}}
\newcommand{\ben}{\begin{enumerate}}
\newcommand{\een}{\end{enumerate}}
\newcommand{\bde}{\begin{widetext}}
\newcommand{\ede}{\end{widetext}}
\newcommand{\nn}{\nonumber}
\newcommand{\crn}{\nonumber \\}
\newcommand{\Tr}{\mathrm{Tr}}
\newcommand{\non}{\nonumber}
\newcommand{\noi}{\noindent}
\newcommand{\al}{\alpha}
\newcommand{\la}{\lambda}
\newcommand{\bet}{\beta}
\newcommand{\ga}{\gamma}
\newcommand{\va}{\varphi}
\newcommand{\om}{\omega}
\newcommand{\pa}{\partial}
\newcommand{\+}{\dagger}
\newcommand{\fr}{\frac}
\newcommand{\sq}{\sqrt}
\newcommand{\bc}{\begin{center}}
\newcommand{\ec}{\end{center}}
\newcommand{\Ga}{\Gamma}
\newcommand{\de}{\delta}
\newcommand{\De}{\Delta}
\newcommand{\ep}{\epsilon}
\newcommand{\varep}{\varepsilon}
\newcommand{\ka}{\kappa}
\newcommand{\La}{\Lambda}
\newcommand{\si}{\sigma}
\newcommand{\Si}{\Sigma}
\newcommand{\ta}{\tau}
\newcommand{\up}{\upsilon}
\newcommand{\Up}{\Upsilon}
\newcommand{\ze}{\zeta}
\newcommand{\ps}{\psi}
\newcommand{\Ps}{\Psi}
\newcommand{\ph}{\phi}
\newcommand{\vph}{\varphi}
\newcommand{\Ph}{\Phi}
\newcommand{\Om}{\Omega}

\newcommand{\Vien}[1]{{\color{red}#1}}
\newcommand{\lh}[1]{{\color{blue}#1}}
\newcommand{\ac}[1]{{\color{cyan}#1}}
\newcommand{\Green}[1]{{\color{green}#1}}
\newcommand{\JM}[1]{{\color{teal}#1}}
\newcommand{\Red}[1]{{\color{red}#1}}
\newcommand{\Revised}[1]{{\color{blue}#1}}
\newcommand{\Blue}[1]{{\color{blue}#1}}
\newcommand{\Long}[1]{{\color{cyan}#1}}
\newcommand{\Cyan}[1]{{\color{cyan}#1}}
\newcommand{\Vienn}[1]{{\color{cyan}#1}}
\title{Fermion masses and mixings and $g-2$ muon anomaly in a $Q_6$ flavored 2HDM}

\author{V. V. Vien}
\email{vvvien@ttn.edu.vn}
\affiliation{Department of Physics, Tay Nguyen University, 
Daklak province, Vietnam.}
\author{H. N. Long}
		\email{hoangngoclong@vlu.edu.vn\, (corresponding author)}	
\affiliation{Subatomic Physics Research Group, Science and Technology Advanced Institute, Van Lang University, Ho Chi Minh City, Vietnam,}
\affiliation{Faculty of Applied Technology, School of
 Technology, Van Lang University, Ho Chi Minh City, Vietnam}
\author{A. E. C\'arcamo Hern\'andez}
\email{antonio.carcamo@usm.cl}
\affiliation{Departamento de Física, Universidad T\'{e}cnica Federico Santa Mar\'{\i}a, Casilla 110-V, Valpara\'{\i}so, Chile.}
\affiliation{Centro Cient\'{\i}fico-Tecnol\'ogico de Valpara\'{\i}so, Casilla 110-V, Valpara\'{\i}so, Chile,}
\affiliation{Millennium Institute for Subatomic physics at high energy frontier - SAPHIR, Fernandez Concha 700, Santiago, Chile.}
\author{Juan Marchant Gonz\'alez}
\email{juan.marchant@upla.cl}
\affiliation{Departamento de Física, Universidad T\'{e}cnica Federico Santa Mar\'{\i}a, Casilla 110-V, Valpara\'{\i}so, Chile.}
\affiliation{Millennium Institute for Subatomic physics at high energy frontier - SAPHIR, Fernandez Concha 700, Santiago, Chile.}
\affiliation{Laboratorio de C\'omputo de F\'isica (LCF-UPLA), Facultad de Ciencias Naturales y Exactas, Universidad de Playa Ancha, Subida Leopoldo Carvallo 270, Valpara\'iso, Chile.}
\date{\today}
\begin{abstract}
We propose an extended 2HDM with $Q_6\times Z_4\times Z_2$ symmetry that can
successfully accommodate the SM fermion mass and mixing hierarchy. The tiny masses of the active neutrinos are generated from a type-I seesaw mechanism mediated by very heavy  right- handed Majorana neutrinos.
  The model gives a natural explanation of the charged lepton mass hierarchy.
Besides that, the experimental values of the physical observables of the neutrino sector: the neutrino mass squared splittings, the leptonic mixing angles and the leptonic Dirac CP violating phase, are also successfully
 reproduced for both normal and inverted neutrino mass hierarchies. We find a feasible range of values
  for the leptonic Dirac CP phase to be in the ranges $\delta_{CP}\in (305.90, 348.70)^\circ$ for normal ordering and $\delta_{CP}\in (308.00, 348.00)^\circ$ for inverted ordering, which is consistent with the 3$\sigma$ experimentally allowed limits.
 The sum of neutrino masses is obtained as $\sum m_i \in (58.03, 60.51)$ meV for normal ordering and $\sum m_i\in (98.07, 101.40)$ meV for inverted ordering which are well consistent with all the recent limits. In addition, the obtained ranges for the effective neutrino masses are $\langle m_{ee}\rangle \in (3.80, 4.38)$\, meV, $m_{\beta} \in (8.53, 9.34)\, \mbox{meV}$ for normal ordering and $\langle m_{ee}\rangle \in (47.85, 49.58) $ meV, $m_{\beta} \in (48.39, 50.09)\, \mbox{meV}$ for inverted ordering which are in agreement with the recent experimental bounds. For the quark sector, the derived results are also in agreement with the recent data on the quark masses and mixing angles. The model under consideration can also accommodate the muon anomalous magnetic moment.
\end{abstract}
\maketitle
\section{\label{intro} Introduction}
The lepton mass hierarchy problem and the observed lepton mixing pattern with the Dirac CP phases are two core issues that imply  
the Standard Model (SM) has to be extended \cite{Feruglio15, Novichkov21}. The unresolved issues within the framework of the SM related to the origins of
(i) the charged-lepton mass hierarchy \cite{PDG2022}, (ii) the neutrino mass hierarchy \cite{Salas2021}, and
(iii) the fermion mixing patterns \cite{PDG2022, Salas2021}.

Non-Abelian discrete symmetries have revealed the advantages in addressing 
the flavour problem (see, for instance, Refs. \cite{King:2013eh,Altarelli:2010gt, King:2015aea, VLA2020S3}). Among the non-Abelian discrete symmetries \cite{Ishi}, $Q_6$ symmetry\footnote{There are some different notations for the irreducible representations of $Q_6$ \cite{Q61,Q62,Q63,Q64,Q65,Q66,Q67, Q68}. We use the notation in Ref. \cite{Ishi}, whereby, four singlets and two doublets of $Q_6$ are denoted by $\mathbf{1}_{++}$, $\mathbf{1}_{--}$, $\mathbf{1}_{+-}$, $\mathbf{1}_{-+}$, and $\mathbf{2}_1, \mathbf{2}_2$.} has attracted the attention because it provides a predictive description of the observed fermion mass and mixing pattern \cite{Q61,Q62,Q63,Q64,Q65,Q66,Q67,Q68,Bonilla:2021ize}. In Ref. \cite{Q61} the symmetry of the SM is supplemented by the $Q_6\times Z_{12R}$ symmetry in which the first two families of left-handed leptons and left-handed quarks are grouped in the $2_1$ of $Q_6$ and the third ones
are assigned as $1_{--}$ while the first two families of  right-handed charged-leptons, right-handed neutrinos and up-and down-right handed quarks are grouped in $2_2$ and the corresponding third ones are assigned as $1_{-+}$ of $Q_6$ with six $SU(2)_L$ doublets and seven singlets are introduced \cite{Q61}. In Ref. \cite{Q62}, the SM is supplemented by the $Q_6\times Z_{4}\times R$ symmetry\footnote{$R$ is the $R$-parity.} in which the first two families of left-handed leptons, right-handed charged leptons, of right-handed neutrinos, and of up-and down-right handed quarks are grouped in $2_2$, the left-handed quarks are grouped in $2_1$ while the third family of left-handed lepton and right-handed charged lepton are put in $1_{++}$, of right-handed neutrino is in $1_{-+}$, of left-handed quark is in $1_{--}$, and of up-and down-right handed quarks are in $1_{+-}$. Further, six doublets and five singlets are needed for generating the fermion masses and mixings. However, the obtained value of $\theta_{13}$ is too small compared to the experimental data \cite{Q62,Salas2021}. In Ref. \cite{Q63}, the SM is supplemented by the $Q_6\times R$ symmetry in which the two first families of left-handed leptons and right-handed charged leptons, of right-handed neutrinos and of up-and down-right handed quarks are grouped in the $2_2$ while the two first families of left-handed quarks are in $2_1$. For the third family, left-and right-handed leptons are put in $1_{++}$, the left-handed quark is in $1_{--}$, the up-and down-right-handed quarks are in $1_{+-}$ and the right-handed neutrino is in $1_{-+}$. Furthermore, six $SU(2)_L$ doublets are needed for generating fermion masses and mixings at the one loop level. However, the obtained value of $\theta_{13}$ is too small compared to the experimental data \cite{Q63,Salas2021}. In Ref. \cite{Q64}, the SM is supplemented by the $Q_6\times Z_4$ symmetry in which the first two families of left-handed leptons and right-handed charged leptons, of right-handed neutrinos and of up-and down-right-handed quarks are grouped in the $2_2$, of left-handed quarks are grouped in the $2_1$ while the third families of left-handed lepton and right-handed charged lepton are put in $1_{++}$, third family of left-handed quark is put in $1_{--}$, the third family of right-handed neutrino is in $1_{+-}$, and the third families of up-and down-right handed quarks are put in $1_{-+}$. Besides, six $SU(2)_L$ doublets and seven singlets are needed for generating fermion masses and mixings with just mentioned $\theta_{12}$ and $\theta_{13}$ without mentioning $\theta_{23}$. Furthermore, the Dirac CP phase is predicted to be beyond $3\sigma$ range for the present experiment data \cite{Q64,Salas2021}. In Ref. \cite{Q65}, the SM is supplemented by the $Q_6\times Z_3\times Z_2$ symmetry in which the first family of left-handed charged lepton is put in $1_{++}$, of  right-handed charged lepton is in $1_{--}$ and of right-handed neutrino is in $1_{+-}$ while the last two families of left-handed charged leptons and right-handed neutrinos are grouped in $2_2$ and of right-handed charged leptons are in $2_1$. On the other hand, one $SU(2)$ doublet and five singlets are needed for generating neutrino masses and mixings with respect to the normal mass hierarchy without mentioning the quark sector \cite{Q65}. In Ref. \cite{Q66}, the SM is supplemented by the $Q_6\times Z_4\times R$ symmetry in which the two first families of left-handed leptons and right-handed charged leptons, right-handed neutrinos and up-and down-right handed quarks are grouped in the $2_2$ and two first families of left-handed quarks are grouped in the $2_1$ while the third family of the left-handed lepton and right-handed charged lepton are put in $1_{++}$, of the left-handed quark is in the $1_{--}$, of right-handed neutrino is in $1_{+-}$ and the third families of up-and down-right handed quark are put in $1_{-+}$. Simultaneously, six $SU(2)_L$ doublets and seven singlets are needed for generating fermion masses and mixings. In Ref. \cite{Q67}, the SM is supplemented by the $Q_6$ symmetry in the framework of CP parities in which the two first families of left-handed lepton,  right-handed neutrino and the third family of left-handed quark and the third family of up right-handed quark are put in $1_{+-}$; the first family of  right-handed charged lepton and the third family of down  right-handed quark are put in $1_{-+}$ while the two last families of left-and right-handed charged leptons,  right-handed neutrinos, the two first families of left-handed quarks and up-and down-right handed quarks are grouped in $2_2$. In addition, six $SU(2)_L$ doublets and one singlet are needed for generating fermion masses and mixings. Only inverted and degenerate hierarchies are considered in which the inverted hierarchy (IH) is less favored than the degenerate one \cite{Q67}. However, the recent data \cite{Salas2021} favor to the normal hierarchy (NH) and IH. In Ref. \cite{Q68}, the symmetry of the SM is supplemented by the $U(1)_{B-L}\times Q_6\times Z_{4}$ symmetry in which the first family of left-handed charged-lepton, right-handed lepton and  right-handed neutrino are put in $1_{+-}$ while the first families of left-handed quark, up-and down-right-handed quarks are put in $1_{-+}$, and all the two other families of quarks and leptons are put in $2_1$ of $Q_6$. In Ref.\cite{Q68}, four $SU(2)_L$ doublets and three singlets are needed for explaining the fermion masses and mixings, however, the "23" and "32" elements of the quark mixing matrix are significantly different from the experimental constraints \cite{Q68}.

The above-mentioned discussion shows that most of the previous models with $Q_6$ symmetry worked with a rather complicated scalar sector involving many $SU(2)_L$ Higgs doublets, like for instance, the model proposed in \cite{Q68}, and not to refer the muon anomalous magnetic moment, whose scalar spectrum contains four $SU(2)_L$ scalar doublets participating in all Yukawa interactions. Thus, it is necessary to find another model based on the $Q_6$ symmetry with a more economical scalar sector that allows to yield predictive fermion mass matrices consistent with the observed pattern of SM fermion masses and mixing angles. In this work, we propose an extended Two-Higgs-Doublets model (2HDM) where the scalar sector is enlarged by the inclusion of five electrically neutral singlet scalar fields and the SM fermion sector is augmented by adding three  right- handed Majorana neutrinos. The SM gauge symmetry is supplemented by the $Q_6\times Z_4\times Z_2$ discrete group, which is crucial to yield predictive textures for the fermion sector consistent with the low energy SM fermion flavor data. The inclusion of the singlet scalar fields in the model is required to build the Yukawa interactions invariant under the symmetries to yield predictive fermion mass matrices consistent with the observed pattern of SM fermion masses and mixings. In the considered model, the tiny active neutrino masses are produced by a type I seesaw mechanism mediated by very heavy  right-handed Majorana neutrinos. We use another assignment for the $Q_6$ symmetry, different than the one considered in \cite{Q68}, in which the first family of the left-handed lepton and right-handed charged lepton ($\psi_{1L}, l_{1R}$) and right-handed neutrino ($\nu_{1R}$) are assigned in $1_{--}$, the second and third families of left-handed leptons are grouped in $2_2$ ($\psi_{L}=(\psi_{2L},\psi_{3L})$) while of the right-handed charged leptons and neutrinos are assigned in $2_1$ ($l_{R}=(l_{2R},l_{3R})$ and $\nu_{R}=(\nu_{2R},\nu_{3R})$). For the quark sector, the first family of the left-handed quark ($Q_{1L}$) is assigned in $1_{++}$ while the two others are grouped in $2_2$ ($Q_{L}=(Q_{2L}, Q_{3L})$); the first families of right-handed quarks ($u_{1R}, d_{1R}$) are assigned in $1_{+-}$ while the two others are grouped in $2_1$ ($u_{R}=(u_{2R},u_{3R})$ and $d_{R}=(d_{2R},d_{3R})$). For scalar fields, one $SU(2)_L$ doublet $S$ put in $1_{+-}$, one $SU(2)_L$ singlet $\phi$ put in $2_{1}$ and two singlets $\rho$ and $\eta$ are respectively put in one representations $1_{+-}$ and $1_{-+}$ of $Q_6$ are added to the SM. This is different from previous works \cite{Q61,Q62,Q63,Q64,Q65,Q66,Q67,Q68}. Consequently, the proposed model
yields a predictive pattern of SM fermion masses and mixings consistent with the experimental data. Furthermore, in this work the muon $g-2$ anomaly is addressed, whereas the model of Ref. \cite{Q68} does not analyze such an important phenomenological issue.

The rest of this work is structured as follows. The description of the model is given in section \ref{model}. The fermion masses and mixings together with the numerical analysis is presented in section \ref{fermion}. The implication of the model in the muon anomalous magnetic moment is discussed in section \ref{anomal}. Finally, some conclusions are stated in section \ref{conclusion}. Appendix \ref{app} presents briefly the Clebsch–Gordan coefficients of the $Q_6$ group. Appendix \ref{Preventedterms} provides the Yukawa terms prevented by additional symmetries $Q_6, Z_4$ and $Z_2$. Appendix \ref{Higgs} shows the model scalar potential.
\section{The model\label{model}}

In this work, three discrete symmetries $Q_6, Z_4, Z_2$, three  right-handed neutrinos $\nu_{1R}, \nu_{2R}, \nu_{3R}$, one doublet $S$ and three singlets $\rho, \eta$ and $\phi$ are added to the SM. In addition to the SM symmetries, additional symmetries $Q_6, Z_4$ and $Z_2$ play important roles in preventing the unwanted terms which are listed in Table \ref{PreventedtermsT} of Appendix \ref{Preventedterms}. The total symmetry of the considered model is $\Gamma= \mathrm{G}_{\mathrm{SM}}\times Q_6\times Z_4\times Z_2$ where $\mathrm{G}_{\mathrm{SM}}$ is the gauge symmetry of the SM. The particle content of the model is summarized in Table \ref{partcont}. All lepton and scalar fields are assigned in the singlet $\mathbf{1}$ representation of $SU(3)_C$ and all quark fields are in the triplet $\mathbf{3}$ of $SU(3)_C$. Furthermore, the hypercharge of doublet scalars
is $\frac{1}{2}$, of right-handed neutrinos and singlet scalars is equal to zero, while those of the SM are retained; thus, they are not shown in Table \ref{partcont}.
\begin{table}[ht]
\begin{center}
\caption{Particle content of the model.}
\vspace{0.2cm}
\begin{tabular}{|c|c|c|c|c|c|c|c|c|c|c|c|c||c|c|c|c|c|c|c|}
\hline
     &$\psi_{1L}$ &$\psi_{L}$ & $l_{1R}$ &$l_{R}$ & $\nu_{1 R}$& $\nu_{R}$ &$Q_{1L}$ & $Q_{L}$ & $u_{1R}$ & $u_{R}$ & $d_{1 R}$& $d_{R}$ &$H$ & $S$  &$\rho$& $\eta$&$\phi$\\ \hline
$Q_6$ & $1_{--}$ & $2_2$&$1_{--}$   & $2_1$ &$1_{--}$&$2_1$&$1_{++}$ & $2_2$&$1_{+-}$   & $2_1$ &$1_{+-}$&$2_1$&$1_{--}$&$1_{+-}$ &$1_{+-}$&$1_{-+}$&$2_1$ \\
$Z_4$&$1$&$1$&$-1$&$i$& $i$&$i$&$i$&$i$&$1$&$1$& $1$&$1$&$1$&$i$&$-i$&$-i$&$-i$\\
$Z_2$&$+$&$-$&$+$&$-$& $-$&$+$&$+$&$+$&$+$&$+$& $+$&$+$&$-$&$-$&$-$&$-$&$-$ \\\hline
 \end{tabular} \label{partcont}
\end{center}
\vspace{-0.25cm}
\end{table}\\
With the particle content given in the Table \ref{partcont}, the scalar doublets $H, S$ and the scalar singlets $\rho, \eta,\phi$ are responsible for generating the charged-lepton mass hierarchy and the left-handed charged-lepton
mixing matrix $\mathbf{U}_{lL}$ which provides the contribution to the lepton mixing matrix of the form $\mathbf{U}=\mathbf{U}^\+_{lL} \mathbf{U}_{\nu}$ with $\mathbf{U}_{\nu}$ being the leptonic mixing matrix. The scalar doublets $H, S$ and the scalar singlet $\phi$ are responsible for generating the Dirac neutrino mass terms whereas the scalar singlets $\rho, \eta, \phi$ yield 
the Majorana neutrino mass terms. Moreover, the scalar doublet $H$ and scalar singlets $\rho, \eta, \phi$ are responsible for generating the masses of the up-and-down quarks. The Yukawa terms invariant under $\Gamma$ read\footnote{Note that $(\bar{\nu}^c_{R}\nu_{R})_{1_{++}}\big(\rho^2+\eta^2+\phi^2\big)_{1_{++}}=0$ due to the antisymmetric form in $\nu^c_{2 R} \nu_{3R}$ and $\nu^c_{3 R} \nu_{2R}$ and $(\phi^2)_{1_{++}}=0$ due to the VEV structure of $\phi$ and the antisymmetric form in $\phi_1 \phi_2$ and $\phi_2 \phi_1$ as a consequence of tensor product $2_1\times 2_1$ of $Q_6$. Furthermore, there exist the other contributions via higher dimensional Weinberg operators,
	$\fr{1}{\La^{2(i+j)+1}}\overline{\psi}_{1L} \psi^c_{1L} \widetilde{\mathbf{H}}^2(\mathbf{H}^\+\mathbf{H})^i (\Phi^* \Phi)^j$ and $\fr{1}{\La^{2(i+j)+1}}\overline{\psi}_{L} \psi^c_{L} \widetilde{\mathbf{H}}^2(\mathbf{H}^\+\mathbf{H})^i (\Phi^* \Phi)^j$ where $\mathbf{H}=H,S$ and $\Phi=\rho, \eta, \phi$	with $i, j = 0, 1, 2, ...$. Since $v_{{\mathbf{H}}}$
and $v_{\mathbf{\Phi}}$ are far smaller than the cut-off scale, i.e., $v_{\mathbf{H}}\ll v_\Phi \ll \La$.
Thus, the left-handed neutrino mass generated via Type II seesaw mechanism
 $\sim v_{\mathbf{H}} \left(\fr{ v_{\mathbf{H}} }{\La}\right)^{2i+1} \left(\fr{v_\mathbf{\Phi}}{\La}\right)^{2j}$ is very small
 compared to the one generated via the type-I seesaw mechanism in Eq. (\ref{Meffv}) and therefore would be neglected.}:
\bea
-\mathcal{L}^{(l)}_{Y}&=&\fr{h_1}{\Lambda} (\bar{\psi}_{1 L}  l_{1 R})_{1_{++}} (S\eta^*)_{1_{++}}
+ \fr{h_2}{\Lambda}(\bar{\psi}_{L} l_{R})_{1_{+-}} (H\eta)_{1_{+-}}
+ \fr{h_3}{\Lambda}(\bar{\psi}_{L} l_{R})_{1_{-+}} (H\rho)_{1_{-+}}\crn
&+&\fr{h_4}{\Lambda}(\bar{\psi}_{L} l_{R})_{2_1} (H \phi)_{2_1}
+ \fr{x_1}{\Lambda} \big[(\bar{\psi}_{1 L} \nu_{R})_{2_{1}}(\widetilde{H}\phi)_{2_{1}}\big]_{1_{++}}
 + x_2 (\bar{\psi}_{L} \nu_{R})_{1_{-+}} \widetilde{S} \crn
 &+& \fr{y_1}{2\Lambda}(\bar{\nu}^c_{1R}\nu_{1R})_{1_{++}}(\rho^2+\eta^2)_{1_{++}}
 + \fr{y_2}{2\Lambda}\left(\bar{\nu}^c_{R}\nu_{R}\right)_{1_{--}} (\rho\eta+\phi^2)_{1_{--}}
+ \fr{y_3}{2\Lambda}\left(\bar{\nu}^c_{R}\nu_{R}\right)_{2_2} (\rho\phi +\eta\phi+\phi^2)_{2_2} +\mathrm{h.c.,}\label{Llep} \eea
 \bea
 -\mathcal{L}^{(q)}_{Y}&=&\fr{h_{1u}}{\Lambda} (\bar{Q}_{1 L}  u_{1 R})_{1_{+-}} (\widetilde{H}\rho^*)_{1_{+-}}
+ \fr{h_{2u}}{\Lambda} (\bar{Q}_{L} u_{R})_{1_{+-}} (\widetilde{H}\rho^*)_{1_{+-}}
+ \fr{h_{3u}}{\Lambda}(\bar{Q}_{L} u_{R})_{1_{-+}} (\widetilde{H} \eta^*)_{1_{-+}}\crn
&+&\fr{h_{4u}}{\Lambda}(\bar{Q}_{L} u_{R})_{2_1} (\widetilde{H} \phi^*)_{2_1}
+ \fr{h_{5u}}{\Lambda}\big[\left(\bar{Q}_{1 L} u_{R}\right)_{2_1} (\widetilde{H} \phi^*)_{2_1} \big]_{1_{++}}
+\fr{h_{6u}}{\Lambda}\big[\left(\bar{Q}_{L} u_{1R} \right)_{2_1} (\widetilde{H} \phi^*)_{2_1} \big]_{1_{++}} \hspace{0.75 cm}\crn
 &+& \fr{h_{1d}}{\Lambda} (\bar{Q}_{1 L}  d_{1 R})_{1_{+-}} (H\rho^*)_{1_{+-}}
+ \fr{h_{2d}}{\Lambda} (\bar{Q}_{L} d_{R})_{1_{+-}} (H \rho^*)_{1_{+-}}
+ \fr{h_{3d}}{\Lambda}(\bar{Q}_{L} d_{R})_{1_{-+}} (H \eta^*)_{1_{-+}} \label{Lquark}\\
&+&\fr{h_{4d}}{\Lambda} (\bar{Q}_{L} d_{R})_{2_1} (H \phi^*)_{2_1}
+ \fr{h_{5d}}{\Lambda}\big[\left(\bar{Q}_{1 L} d_{R}\right)_{2_1}(H \phi^*)_{2_1}\big]_{1_{++}}
+\fr{h_{6d}}{\Lambda}\big[\left(\bar{Q}_{L} d_{1R}\right)_{2_1}(H \phi^*)_{2_1}\big]_{1_{++}}  \crn
&+& \mathrm{h.c,}\nonumber
\eea
where $h_{i}$,  $x_{j}$, $y_{k}$, $h_{lu}$ and $h_{ld}$\, $(i=1,\dotsc, 4; j=1, 2; k=1,\dotsc, 3;
l=1,\dotsc, 6)$ are Yukawa-like couplings, and $\Lambda$ is the cut-off scale.

From the charged leptonic Yukawa terms, it follows that there are no tree level lepton flavor-changing scalar interactions. This is due to the fact that despite there are two $SU(2)$ scalar doublets in the leptonic Yukawa interactions, one of them, i.e., $S$ only interacts with the first family of the SM charged leptonic fields, whereas the remaining one, i.e., $H$ exhibits Yukawa interactions with the second and third family of the SM charged leptons. Furthermore, only one $SU(2)$ scalar doublet participates in the quark Yukawa interactions, then implying the absence of tree level flavor changing neutral processes in the quark sector.

As will be presented in Appendix \ref{Higgs}, the vacuum expectation values (VEVs) of scalar fields take the following structures:
\bea
&&\langle H \rangle = (0 \hs \hs v_H)^T,\hs \langle S \rangle = (0 \hs \hs v_S)^T, \hs \langle \rho \rangle = v_\rho, \hs \langle \eta \rangle = v_\eta,\hs \langle \phi \rangle =\left(v_\phi, \hs v_\phi\right). \label{scalarvev}
\eea
In fact, the electroweak symmetry breaking scale is about one hundred GeV while the cut-off scale is still unknown and is assumed to be very high with $\Lambda\in (10^{13},\, 10^{15})\, \mathrm{GeV}$ \cite{Pokorski05}, $\Lambda\in (10^{5},\, 10^{19})\, \mathrm{GeV}$ \cite{Chunjhep12}, and $\Lambda$ is about $10^{13}\, \mathrm{GeV}$ \cite{cutoffscal21}. In the model under consideration, we assume for the sake of simplicity that the VEVs of the scalar singlets are of the same order of magnitude and we consider a particular benchmark scenario where the model cut-off is taken to be \footnote{Our benchmark choice $\Lambda \simeq 10^{13}$ GeV for the model cut-off is motivated by the analysis made in Ref. \cite{cutoffscal21} where the cut-off scale of the SM can be lowered to $\Lambda \simeq 10^{13}$ GeV by applying gravitational positivity bounds to the light-by-light scattering by considering the electroweak theory without the QCD sector.}
$\Lambda \simeq 10^{13}$ GeV,
\bea
&&v_H^2+v_S^2 =246^2 \, \mathrm{GeV}^2, \hs \Lambda \simeq 10^{13}\, \mathrm{GeV}\Revised{,} \label{LaSMscale}\\
&&v_\rho \simeq v_\eta  \simeq v_\phi = 5\times 10^{11} \, \mathrm{GeV}. \label{vevscales}
\eea
On the other hand, the charged-lepton mass hierarchy requires that $v_S$ to be less than $v_H$ about two orders of magnitude, $v_S\sim 10^{-2} v_H$, thus we obtain
\bea
&&v_S\simeq 2.46\, \mathrm{GeV}, \hs v_H\simeq 246.00\, \mathrm{GeV}. \label{vSvH}
\eea
It is worth mentioning that our choice $\Lambda \simeq 10^{13}$ GeV corresponds to a specific benchmark scenario and does not necessarily imply that New Physics should appear at that scale. The scale $\Lambda$ corresponds to the energy scale where our model has to be replaced by a more fundamental theory, like for instance a GUT. Note that our model has non SM scalars with masses at the subTeV scale, whose production at colliders will be an indication of New Physics.

In 2HDM, two scalar doublets $H=\left(\zeta^{\+}_{H}, \hspace{0.15 cm} v_H +\sigma_H+i\eta_H \right)^T, \, S=\left(\zeta^{\+}_{S}, \hspace{0.15 cm} v_S+\sigma_S+i\eta_S \right)^T$ contain eight fields including $v_H, v_S, \zeta^{\+}_{H}, \zeta^{\+}_{S}, \sigma_{H}, \sigma_{S}, \eta_{H}$ and $\eta_{S}$ where three of them 
are eaten to give mass to the gauge bosons $W^{\pm}$ and $Z^0$. The remaining
five are physical Higgs states including one CP-even neutral Higgs ($H^0$), one CP-odd neutral Higgs ($A^0$), the SM-like Higgs boson ($h$), and a pair of charged states ($H^{\pm}$).
The masses of the gauge particles and the Higgs particles are in consistent with the experimental requirements which has been considered in various previous works (for more details, see for instance \cite{Wang2hdm,Cao2hdm,Chun2hdm,Chun2hdm15,Abe2hdm}). On the other hand, the
scalar potential includes mixings among all the doublet and singlet scalars, however, the doublet-singlet mixings and singlet-singlet mixings are very small compared to that of doublets. Therefore, the couplings between doublets $\{H, S\}$ and singlets $\{\phi, \rho, \eta\}$ and the mixing between singlet scalars will be negligible. In this sense, the total $V_{\mathrm{scalar}}$ 
is reduced to $V(H)+V(S)+ V(H,S)$ which is similar to that of Ref. [23] but in our model two couplings $(H^{\dagger} H)(S^{\dagger} S)$ and $(H^{\dagger} S)(S^{\dagger} H)$ are prevented by $Q_6$ symmetry. Furthermore, equation (\ref{vSvH}) implies that $\tan \beta= \frac{v_H}{v_S}$ = 100 which is of the same order as those of Refs. \cite{Wang2hdm,Cao2hdm,Chun2hdm,Chun2hdm15,Abe2hdm,Crivellin:2015hha} with $t_\beta \in (32, 80)$ \cite{Wang2hdm}, $t_\beta \in (37, 80)$ \cite{Cao2hdm}, $t_\beta \in (30, 100)$ \cite{Chun2hdm}, $t_\beta \in (60, 100)$ \cite{Crivellin:2015hha}, $t_\beta \in (10, 150)$ \cite{Chun2hdm15} and $t_\beta \sim 1000$ \cite{Abe2hdm}\Revised{.}

\section{\label{fermion}Fermion masses and mixings}
\subsection{\label{lep}Lepton masses and mixings}
Using the tensor product of $Q_6$ symmetry \cite{Ishi}, after symmetry breaking, from Eqs. (\ref{Llep}) and (\ref{scalarvev}), we obtain the mass matrix for the charged-leptons as follows:
 \bea M_l&=&\left(%
\begin{array}{ccc}
  a_l& 0      & 0 \\
  0  & b_l+c_l& -d_l\\
  0  & -d_l  & -b_l+c_l\\
\end{array}%
\right), \label{Mclep}\eea
where
\bea
&&a_l=\fr{h_1}{\Lambda} v_S v_\eta, \hs
 b_l = \fr{h_2}{\Lambda} v_H v_\eta, \hs
 c_l = \fr{h_3}{\Lambda} v_H v_\rho, \hs d_l=\fr{h_4}{\Lambda} v_H v_\phi. \label{abcdl}
\eea
The Yukawa couplings $h_{i}$ are complex; thus, $a_l, b_l, c_l$ and $d_l$ are complex too, and $M_l$ in (\ref{Mclep}) is a complex matrix. For simplicity, we consider the case of $\arg b_l=\arg c_l$. Let us define a Hermitian matrix, $\mathbf{m}^2_l=  M_l M^+_l$, given by
\bea
\mathbf{m}^2_l&=& \left(
\begin{array}{ccc}
 |a_{l}|^2 & 0 & 0 \\
 0 & \big(|b_{l}|+|c_{l}|\big)^2+|d_{l}|^2 &\hs -2 |d_{l}| \sqrt{|b_{l}|^2 s_\kappa^2+|c_{l}|^2 c_\kappa^2}. e^{-i\theta} \\
 0 & -2 |d_{l}| \sqrt{|b_{l}|^2 s_\kappa^2+|c_{l}|^2 c_\kappa^2}. e^{i\theta} & \big(|b_{l}|-|c_{l}|\big)^2+|d_{l}|^2 \\
\end{array}
\right), \label{mlsq}
\eea
where\footnote{In this work we use the following notations: $s_\psi=\sin \psi,\, c_\psi=\cos \psi, \, s_\theta=\sin \theta,\, c_\theta=\cos \theta,\, t_{\kappa}=\tan \kappa,\,t_{\theta}=\tan\theta,\, s_\delta=\sin\delta_{CP},\, s_{ij}=\sin\theta_{ij},\, c_{ij}=\cos\theta_{ij}$ and $t_{ij}=\tan\theta_{ij}\, (ij=12,13,23)$.}
\bea
\kappa=\arg b_l-\arg d_l, \hs t_\theta =\frac{|b_l|}{|c_l|} t_\kappa. \label{altheta}
\eea
The matrix $\mathbf{m}^2_l$ in Eq. (\ref{mlsq}) is diagonalised by the unitary matrix $\mathbf{U}_{lL}$, satisfying $\mathbf{U}^+_{lL} \mathbf{m}^2_l \mathbf{U}_{lL}=\mathrm{diag} (m^2_e, m^2_\mu, m^2_\tau)$, where
\bea
&&m^2_e =|a_{l}|^2, \,\, m^2_{\mu, \tau} =|b_{l}|^2+|c_{l}|^2+|d_{l}|^2 \mp 2 |b_{l}||c_{l}| \sqrt{\frac{|b_{l}|^2+t_\theta^2 \left(|c_{l}|^2+|d_{l}|^2\right)+|d_{l}|^2}{|b_{l}|^2+|c_{l}|^2 t_\theta^2}}, \label{memt}\\
&&\mathbf{U}_{lL}=
\left(%
\begin{array}{ccc}
  1 & 0 & 0 \\
   0 & c_\psi &\,\,\,\,\,\, -s_\psi . e^{-i \theta}\\
  0 &\hs s_\psi . e^{i \theta} & c_\psi  \\
\end{array}%
\right), \label{UClep}\\
&&s_\psi=\frac{1}{\sqrt{\frac{|d_{l}|^2 \left(t_\theta^2+1\right)}{\left(|b_{l}|^2+|c_{l}|^2 t_\theta^2\right) \left(\sqrt{\frac{|b_{l}|^2+t_\theta^2 \left(|c_{l}|^2+|d_{l}|^2\right)+|d_{l}|^2}{|b_{l}|^2+|c_{l}|^2 t_\theta^2}}+1\right)^2}+1}}. \label{spsi} \eea
Eqs.  (\ref{abcdl}), (\ref{memt}) and (\ref{spsi}) yield the following relations,
\bea
&&|h_1|=\frac{\Lambda m_e}{v_S v_\eta}, \hs
|h_2|=\frac{\Lambda}{2{v_H v_\eta}} \sqrt{\frac{\big(m_\mu^2+m_\tau^2\big) c_{2\psi}^2 \left(t_\theta^2+1\right)+2 \Delta_h}{c_{2\psi}^2 t_\theta^2+1}}, \label{h1h2}\\
&&|h_3|=\frac{\Lambda}{2v_H v_\rho} \sqrt{\frac{\big(m_\mu^2+m_\tau^2\big) c_{2\psi}^2 \left(t_\theta^2+1\right)-2 \Delta_h}{c_{2\psi}^2+t_\theta^2}}, \label{h3}\\
&&|h_4|=\frac{\Lambda}{v_H v_\phi} \sqrt{\frac{s_\psi^2 c_\psi^2 \left\{-2 \big(t_\theta^2-1\big) \Delta_h+\big(m_\mu^2+m_\tau^2\big) \big[(t_\theta^2+1)^2-4 s_\psi^2 c_\psi^2 (t_\theta^4+1)\big]\right\}}{\left(c_{2\psi}^2+t_\theta^2\right) \left(c_{2\psi}^2 t_\theta^2+1\right)}}, \label{h4}
\eea
where
\bea
\Delta_h=\sqrt{ m_\mu^2 m_\tau^2 c_{2\psi}^4 \big(t_\theta^4+1\big)+t_\theta^2 \left[2 m_\tau^2 m_\mu^2 c_{2\psi}^4-4 s_\psi^4 \big(2 s_\psi^4-3 s_\psi^2+1\big)^2 \big(m_t^2-m_\mu^2\big)^2\right]}. \label{Delh}
\eea
Equations (\ref{h1h2})-(\ref{Delh}) and (\ref{cthetaexpres}) show that four parameters $h_{1}, h_{2}, h_{3}$ and $h_{4}$ are expressed in terms of scalar VEVs, cut-off scale $\Lambda$, three charged lepton masses $m_e, m_\mu, m_\tau$, three neutrino mixing angles $\theta_{ij}\, (ij=12,13,23)$ and one 
parameter $\psi$ which plays a crucial role in reproducing the range of the Dirac CP violating phase $\delta_{CP}$.

We now come back to the neutrino sector. With the aid of the tensor product of $Q_6$ symmetry \cite{Ishi}, Eqs. (\ref{Llep}) and (\ref{scalarvev}), after symmetry breaking, we obtain the $6\times 6$ neutrino mass
matrix in the basis $(\nu^c_L,\, \nu_R)$,
\bea \mathbf{M}&=&\left(%
\begin{array}{ccc}
0 & \mathrm{M}_{D} \\
\mathrm{M}^T_D & \mathrm{M}_{R}\\
\end{array}%
\right), \label{Mnu66}\eea
where $\mathrm{M}_{D}$ and $\mathrm{M}_R$ are Dirac and Majorana neutrino mass matrices having the forms of
\bea \mathrm{M}_D&=&\left(%
\begin{array}{ccc}
0 &\hs\,\, -a_{D} & a_{D} \\
0 &\hs\,\, b_{D} & 0 \\
0 &\hs 0       &b_{D} \\
\end{array}%
\right),
\hs \mathrm{M}_R=
\left(%
\begin{array}{ccc}
a_R&\hs\, 0 &\hs\,\, 0 \\
0  &\hs\,\, c_R-d_R &\hs\,\, b_R\\
0  &\hs\,\,  b_R  & \hs c_R+d_R \\
\end{array}%
\right),  \label{MDR}\eea
with
\bea
&&a_D=\fr{x_1}{\Lambda} v_H v_\phi,\,\, b_D =x_2 v_S, \crn
&&a_R =\fr{y_1}{\Lambda} \big(v^2_\rho+v^2_\eta\big), \,\,
b_R =\fr{y_2}{\Lambda}\left(v_\rho v_\eta
 + 2v_\phi^2\right),\hs c_R= \fr{y_3}{\Lambda} v_\rho v_\phi, \hs d_R=\fr{y_3}{\Lambda} (v_\eta+v_\phi)v_\phi. \label{abcDR}\eea
The mass matrix $\mathbf{M}$ in Eq. (\ref{Mnu66}) can be diagonalised by a $6\times 6$
matrix $\mathbf{U}_{6\times 6}$  satisfying,  
\begin{equation}\label{DiagonaliseTheMass}
\mathbf{U}^{\dagger}_{6\times 6} \mathbf{M} \mathbf{U}^{\ast}_{6\times 6}=\begin{pmatrix}\mathrm{m}_{\nu}^{\rm diag} & \mathbf{0}\\
\mathbf{0}& \mathrm{M_N}^{\rm diag} \end{pmatrix},
\end{equation}
where
\begin{align}
&\mathbf{U}_{6\times 6}= \begin{pmatrix} c_\vartheta & s_\vartheta\\
-s^\dagger_\vartheta & c^\dagger_\vartheta  \end{pmatrix}
\begin{pmatrix} \mathbf{U}_{\nu} & \mathbf{0}\\
\mathbf{0} & \mathbf{U}_N^{\ast} \end{pmatrix}, \label{U66}\\
&s_\vartheta=\sum_{k=0}^\infty \frac{(-\vartheta \vartheta^\dagger)^k \vartheta}{(2k+1)!}, \hspace{0.25 cm} c_\vartheta=\sum_{k=0}^\infty \frac{(-\vartheta\vartheta^\dagger)^k}{(2k)!},  \label{scvartheta}\end{align}
with $\vartheta =M_D M^{-1}_R$ is the matrix that represents the mixing between the the active neutrinos and the heavy mass
eigenstates, and $\mathbf{U}_{\nu}$ and $\mathbf{U}_{N}$ are the matrices that diagonalize the light neutrino ($m_\nu$) and the heavy neutrino ($\mathrm{M}_N$) mass matrices, respectively,
\begin{align}
&\mathrm{m}_{\nu}^{\rm diag}= \mathbf{U}_{\nu}^{\dagger}m_{\nu} \mathbf{U}_{\nu}^{\ast}=\text{diag}(m_1,m_2,m_3), \label{mlightef}\\
&\mathrm{M_N}^{\rm diag}=\mathbf{U}_N^T \mathrm{M}_N \mathbf{U}_N=\text{diag}(M_1,M_2,M_3)\label{MNdiagDef}.
\end{align}
In the seesaw limit, i.e., $\mathrm{M}_D\ll \mathrm{M}_R$, the parameter $\vartheta$ that indicates the mixing between the active neutrinos $\nu_L$ and the right-handed (sterile) neutrinos $\nu_R$, is given by
\begin{equation}
\vartheta = M_D M_M^{-1} \ll \mathbf{1}_{3\times3}. \label{vartheta}
\end{equation}
Equations (\ref{scvartheta}) and (\ref{vartheta}) yield the first approximation of $\vartheta$,
\begin{align}
&s_\vartheta\simeq \vartheta, \hspace{0.25 cm} s^{\dagger}_\vartheta\simeq \vartheta^{\dagger}, \hspace{0.25 cm}
c_\vartheta=\mathbf{1}_{3\times3}-\frac{\vartheta \vartheta^{\dagger}}{2}, \hspace{0.25 cm}
c^{\dagger}_\vartheta=\mathbf{1}_{3\times3}-\frac{\vartheta^{\dagger} \vartheta}{2},\end{align}
and the $6\times 6$ mixing matrix $\mathbf{U}_{6\times 6}$ in Eq. (\ref{U66}) is reduced to
\begin{equation}
\mathbf{U}_{6\times 6} \simeq \begin{pmatrix} \mathbf{1}_{3\times3}-\frac{\vartheta \vartheta^{\dagger}}{2} & \theta \\
-\theta^{\dagger} & \mathbf{1}_{3\times3}-\frac{\vartheta^{\dagger} \vartheta }{2} \end{pmatrix} \begin{pmatrix} \mathbf{U}_{\nu} & \mathbf{0} \\
\mathbf{0} & \mathbf{U}_N^{\ast} \end{pmatrix}. \label{U66reduced1}
\end{equation}
The light neutrino ($m_\nu$) and the heavy neutrino ($\mathrm{M}_N$) mass matrices obtained as:
\begin{align}
&m_{\nu}\simeq -\mathrm{M}_D{\mathrm{M}}_R^{-1}\mathrm{M}^T_D, \label{mlightseeesaw}\\
&M_N \simeq \mathrm{M}_R+\frac{1}{2}\left( \vartheta^{\dagger} \vartheta \mathrm{M}_R+ \mathrm{M}^T_R \vartheta^T \vartheta^{\ast} \right)\equiv \mathrm{M}_R+ \delta M_R.\label{Mheavy}
\end{align}
In the considered model, equations (\ref{LaSMscale}), (\ref{vevscales}), (\ref{MDR}) and (\ref{abcDR}) imply that\footnote{Here, $v^2_X=v^2_\rho+v^2_\eta \simeq v_\rho v_\eta
 + 2v_\phi^2\simeq v_\rho v_\phi\simeq(v_\eta+v_\phi)v_\phi$, and to estimate the order of magnitude of $\mathrm{M}_D$ and $\mathrm{M}_R$ we consider the Yukawa-like couplings in the neutrino sector to be of the same order of magnitude (and the same order as $x_0$).}
\bea
 &&|\mathrm{M}_D|\simeq 10\, \mathrm{GeV} \ll  10^{10} \mathrm{GeV} \sim \frac{v^2_X}{\Lambda}\simeq |\mathrm{M}_R|,
 \eea
 and $\vartheta$ in Eq. (\ref{vartheta}) is estimated as
 \begin{equation}
\vartheta \simeq 10^{-10}\left(
\begin{array}{ccc}
 0. & 2.46 & -0.82 \\
 0. & -0.246 & 0.246 \\
 0. & 0.246 & 0.082 \\
\end{array}
\right), \hspace{0.25 cm}\delta M_R \sim 10^{-9} x_0\left(
\begin{array}{ccc}
 0. & 0. & 0. \\
 0. & -3.09  & 2.08 \\
 0. & 2.08 & -0.99  \\
\end{array}
\right). \label{vartheta}
\end{equation}
The following comments are in order:
\begin{itemize}
  \item [$(i)$] Since $\delta M_R$ is very small compared to the right-handed neutrino mass matrix $M_R$, we have $M_N\simeq M_R$.
 \item [$(ii)$] Since $\vartheta$ is very small compared to the identity matrix $\mathbf{1}_{3\times 3}$, the mixing
between the active neutrinos $\nu_L$ and the right-handed (sterile) neutrinos $\nu_R$ is negligible.
  \item [$(iii)$] With the aid of Eq. (\ref{vartheta}), the mixing matrix $\mathbf{U}_{6\times 6}$ in Eq. (\ref{U66reduced1}) is reduced to
\begin{equation}
\mathbf{U}_{6\times 6} \simeq \begin{pmatrix} \mathbf{U}_{\nu} & \theta \mathbf{U}^*_{N} \\
 -\theta^{\dagger} \mathbf{U}_{\nu} & \mathbf{U}^*_N \end{pmatrix}\simeq \begin{pmatrix} \mathbf{U}_{\nu} & \mathbf{0} \\
 \mathbf{0} & \mathbf{U}^*_N \end{pmatrix}. \label{U66reduced2}
\end{equation}
The matrices $\mathbf{U}_{\nu}$ and $\mathbf{U}_{N}$ are determined in Sec. \ref{activeneutrino} and Sec. \ref{heavyneutrino}, respectively.
\end{itemize}
\subsubsection{\label{activeneutrino} Light neutrino mass}
The (effective) light neutrino mass matrix $m_\nu$ in Eq. (\ref{mlightseeesaw}) has the following explicit form:
 \bea
m_\nu&=&a\left(
\begin{array}{ccc}
 \frac{2 (b_{R}+c_{R})}{b} & -b_{R}-c_{R}-d_{R} & b_{R}+c_{R}-d_{R} \\
 -b_{R}-c_{R}-d_{R} & b (c_{R}+d_{R}) & -b b_{R} \\
 b_{R}+c_{R}-d_{R} & -b b_{R} & b (c_{R}-d_{R}) \\
\end{array}
\right),\label{Meffv}\eea
where
 \bea
 &&a=\frac{a_{D} b_{D}}{b_{R}^2-c_{R}^2+d_{R}^2}, \hs b=\frac{b_D}{a_D}, \label{ab}\eea
with $a_{D, R}, b_{D,R}$, $c_R$ and $d_R$ are defined in Eq. (\ref{abcDR}).

The Yukawa couplings $x_{j}$ and $y_k$ are complex; thus, $a_{D, R}, b_{D,R}$, $c_R$ and $d_R$ are complex, and $m_\nu$ in Eq. (\ref{Meffv}) is a complex matrix. Furthermore, the expression (\ref{Meffv}) shows that there are five complex parameters in the neutrino sector including $a,\, b, \, b_R$, $c_R$  and $d_R$, thus thre are ten real parameters. Considering the case of real VEVs for the scalar fields, $c_R$ and $d_R$ have the same argument. Furthermore, the phase redefinition of the  right-handed neutrino fields $\nu_{1R}$ and $\nu_{R}$ allows to rotate away the phase of three couplings\footnote{The phases in Yukawa couplings $y_{k}\, (k=1,2,3)$ can be removed by redefining the right-handed neutrinos and the Yukawa couplings through the transformation $\nu_{1R}\rightarrow \nu_{1R} e^{-i\omega_\nu}$, $\nu_{R}\rightarrow \nu_{R} e^{-i\omega_\nu}$ and $y_{k}\rightarrow y_{k} e^{i2 \omega_\nu}$.} $y_{1,2,3}$, i.e, four parameters $a_R,\, b_R, \, c_R$ and $d_R$ are real. Furthermore, one parameter $\alpha_a=\arg (a)$ is absorbed and five parameters $a_0=|a|, b_0=|b|, B_0=|b_R|, \hs C_0=|c_R|$ and $D_0=|d_R|$ are combined into three new parameters $\mathbf{X}, \mathbf{Y}$ and $\mathbf{Z}$ in the process of generating the Hermitian matrix $\mathbf{m}_{\mathrm{eff}}^2 = m_\nu m^{+}_\nu$. Therefore, there exist five real parameters in the neutrino sector, including $\mathbf{X}, \mathbf{Y}, \mathbf{Z}, b_0$ and $\gamma =\arg (b_D)$.
The Hermitian matrix $\mathbf{m}_{\mathrm{eff}}^2$, possesses real and positive eigenvalues, takes the following form:
\bea
&&\mathbf{m}_{\mathrm{eff}}^2=\left(
\begin{array}{ccc}
 \frac{2 \mathbf{X}}{b^2_0} & -\frac{(\mathbf{X}+\mathbf{Y}) e^{-i \gamma}}{b_0} & \frac{(\mathbf{X}-\mathbf{Y}) e^{-i \gamma}}{b_0} \\
-\frac{(\mathbf{X}+\mathbf{Y}) e^{i \gamma}}{b_0} & \mathbf{X}+\mathbf{Y}-\mathbf{Z} & -\mathbf{Z} \\
\frac{(\mathbf{X}-\mathbf{Y}) e^{i \gamma}}{b_0} & -\mathbf{Z} & \mathbf{X}-\mathbf{Y}-\mathbf{Z} \\
\end{array}
\right), \label{Msqv}
 \eea
 where
 \bea
 &&\mathbf{X}=a_{0}^2 \left[\left(b_{0}^2+2\right) (B_{0}+C_{0})^2+b_{0}^2 D_{0}^2\right],\hs
 \mathbf{Y}=2 a_{0}^2 D_{0} \left(b_{0}^2 C_{0}+B_{0}+C_{0}\right), \crn
&&\mathbf{Z}=a_{0}^2 \left[2 \left(b_{0}^2+1\right) B_{0} C_{0}+B_{0}^2+C_{0}^2-D_{0}^2\right]. \label{XYZ}
 \eea
The matrix $\mathbf{m}_{\mathrm{eff}}^2$ is diagonalized by $\mathbf{R}_\nu$ satisfying $\mathbf{R}^{\dagger}_\nu \mathbf{m}_{\mathrm{eff}}^2 \mathbf{R}_\nu =\mathrm{diag}\big(\lambda^2_{1}, \lambda^2_{1}, \lambda^2_{3}\big)$ where
three eigenvalues and the corresponding eigenvectors are given by
\bea && \lambda^2_{1} =0, \hs \lambda^2_{2, 3}=\Omega_1\mp \Omega_2, \label{m123sq}\\
&&\mathbf{R}_\nu=\left(
\begin{array}{ccc}
 -\fr{b_0}{\sqrt{b_0^2+2}} & \fr{\kappa_1}{\sqrt{\kappa_1^2+ \tau_1^2+1}} & \fr{\kappa_2}{\sqrt{\kappa_2^2+\tau_2^2+1}} \\
 -\fr{e^{i \ga}}{\sqrt{b_0^2+2}} & \fr{e^{i \ga} \tau_1}{\sqrt{\kappa_1^2+\tau_1^2+1}} & -\fr{e^{i \ga} \tau_2}{\sqrt{\kappa_2^2+\tau_2^2+1}} \\
 \fr{e^{i \ga}}{\sqrt{b_0^2+2}} & \fr{e^{i \ga}}{\sqrt{\kappa_1^2+\tau_1^2+1}} & \fr{e^{i \ga}}{\sqrt{\kappa_2^2+\tau_2^2+1}} \\
\end{array}
\right),\label{U0}\eea
where
\bea
&&\Omega_1=\left(\frac{1}{b_{0}^2}+1\right)\mathbf{X}-\mathbf{Z}, \hs \Omega_2=\fr{\sqrt{\Omega_0}}{b_0^2}, \hs \Omega_0=b_0^2 \big(2 + b_0^2\big) \mathbf{Y}^2 + \big(\mathbf{X} + b_0^2 \mathbf{Z}\big)^2, \label{Om12}
\eea
with $\mathbf{X,Y, Z}$ are defined in Eq. (\ref{XYZ}), and $\kappa_{1,2}$ and $\tau_{1,2}$ are given by
\bea
&&\kappa_{1,2}=\frac{\mathbf{X}+b_{0}^2 (\mathbf{Y}+\mathbf{Z})\mp \sqrt{\Omega_0}}{b_{0} \left(\mathbf{X}-\mathbf{Y}+b_{0}^2 \mathbf{Z}\right)}, \hs \tau_{1,2}=\frac{\sqrt{\Om_0}\mp \left(b_{0}^2+1\right) \mathbf{Y}}{\mathbf{X}-\mathbf{Y}+b_{0}^2 \mathbf{Z}}, \label{kk1k2n1n2expres}
\eea
which satisfy the following relations:
\bea
&& b_0 \kappa_1+\tau_1-1=0,\hs b_0 \kappa_2 - \tau_2-1=0,\hs \kappa_1 \kappa_2 - \tau_1 \tau_2+1=0. \label{KKiNi}
\eea
It is noted that the lightest neutrino mass eigenvalue $\lambda^2_1=0$ is corresponding to the neutrino eigenstate \\ $\bigg(-\fr{b_0}{\sqrt{b_0^2+2}}, \hs -\fr{e^{i \ga}}{\sqrt{b_0^2+2}}$,$ \hs \fr{e^{i \ga}}{\sqrt{b_0^2+2}}\bigg)^T$ and $\lambda^2_2 < \lambda^2_3$. Therefore, the mass ordering should
be either $\big(0, \lambda^2_2, \lambda^2_3\big)$ or $\big(\lambda^2_2, \lambda^2_3,  0\big)$ and the neutrino mass matrix $\mathbf{m}_{\mathrm{eff}}^2$ in Eq. (\ref{Msqv}) can be diagonalized as
 \begin{equation}
\mathbf{U}_{\nu }^+ \mathbf{m}_{\mathrm{eff}}^2 \mathbf{U}_{\nu }=\left\{
\begin{array}{l}
\left(
\begin{array}{ccc}
0 & 0 & 0 \\
0 & \lambda^2_{2} & 0 \\
0 & 0 & \lambda^2_{3}%
\end{array}%
\right),\hspace{0.1cm} \mathbf{U}_\nu=\left(
\begin{array}{ccc}
 -\fr{b_0}{\sqrt{b_0^2+2}} & \fr{\kappa_1}{\sqrt{\kappa_1^2+ \tau_1^2+1}} & \fr{\kappa_2}{\sqrt{\kappa_2^2+\tau_2^2+1}} \\
 -\fr{e^{i \ga}}{\sqrt{b_0^2+2}} & \fr{e^{i \al} \tau_1}{\sqrt{\kappa_1^2+\tau_1^2+1}} & -\fr{e^{i \al} \tau_2}{\sqrt{\kappa_2^2+\tau_2^2+1}} \\
 \fr{e^{i \ga}}{\sqrt{b_0^2+2}} & \fr{e^{i \al}}{\sqrt{\kappa_1^2+\tau_1^2+1}} & \fr{e^{i \al}}{\sqrt{\kappa_2^2+\tau_2^2+1}} \\
\end{array}
\right) \hspace{0.2cm}\mbox{for NH},\ \  \\
\left(
\begin{array}{ccc}
\lambda^2_{2} & 0 & 0 \\
0 & \lambda^2_{3} & 0 \\
0 & 0 & 0
\end{array}%
\right) ,\hspace{0.1cm} \mathbf{U}_\nu=\left(
\begin{array}{ccc}
 \fr{\kappa_1}{\sqrt{\kappa_1^2+ \tau_1^2+1}} &\fr{\kappa_2}{\sqrt{\kappa_2^2+\tau_2^2+1}}&-\fr{b_0}{\sqrt{b_0^2+2}}  \\
  \fr{e^{i \al} \tau_1}{\sqrt{\kappa_1^2+\tau_1^2+1}} &-\fr{e^{i \al} \tau_2}{\sqrt{\kappa_2^2+\tau_2^2+1}} & -\fr{e^{i \ga}}{\sqrt{b_0^2+2}} \\
 \fr{e^{i \al}}{\sqrt{\kappa_1^2+\tau_1^2+1}} & \fr{e^{i \al}}{\sqrt{\kappa_2^2+\tau_2^2+1}} & \fr{e^{i \ga}}{\sqrt{b_0^2+2}} \\
\end{array}
\right) \hspace{0.2cm}\mbox{for IH.}%
\end{array}%
\right.  \label{Unu}
\end{equation}
The corresponding leptonic mixing matrix, $\mathbf{U}=\mathbf{U}_{lL}^{\dag} \mathbf{U}_{\nu }$, is
\begin{equation}
\mathbf{U}=\left\{
\begin{array}{l}
\left(
\begin{array}{ccc}
 -\frac{b_{0}}{\sqrt{b_{0}^2+2}} & \frac{\kappa_{1}}{\sqrt{\kappa_{1}^2+\tau_{1}^2+1}} & \frac{\kappa_{2}}{\sqrt{\kappa_{2}^2+\tau_{2}^2+1}} \\
 \frac{e^{i \gamma} \left(e^{-i \theta} s_\psi-c_\psi\right)}{\sqrt{b_{0}^2+2}} & \frac{e^{i \gamma} \left(e^{-i \theta} s_\psi+c_\psi \tau_{1}\right)}{\sqrt{\kappa_{1}^2+\tau_{1}^2+1}} & \frac{e^{i \gamma} \left(e^{-i \theta} s_\psi-c_\psi \tau_{2}\right)}{\sqrt{\kappa_{2}^2+\tau_{2}^2+1}} \\
 \frac{e^{i \gamma} \left(c_\psi+e^{i \theta} s_\psi\right)}{\sqrt{b_{0}^2+2}} & \frac{e^{i \gamma} \left(c_\psi-e^{i \theta} s_\psi \tau_{1}\right)}{\sqrt{\kappa_{1}^2+\tau_{1}^2+1}} & \frac{e^{i \gamma} \left(c_\psi+e^{i \theta} s_\psi \tau_{2}\right)}{\sqrt{\kappa_{2}^2+\tau_{2}^2+1}} \\
\end{array}
\right) \hspace{0.2 cm}\mbox{for  NH},  \label{Ulep}  \\
\left(
\begin{array}{ccc}
  \frac{\kappa_{1}}{\sqrt{\kappa_{1}^2+\tau_{1}^2+1}} &\frac{\kappa_{2}}{\sqrt{\kappa_{2}^2+\tau_{2}^2+1}} & -\frac{b_{0}}{\sqrt{b_{0}^2+2}} \\
  \frac{e^{i \gamma} \left(e^{-i \theta} s_\psi+c_\psi \tau_{1}\right)}{\sqrt{\kappa_{1}^2+\tau_{1}^2+1}} &\frac{e^{i \gamma} \left(e^{-i \theta} s_\psi-c_\psi \tau_{2}\right)}{\sqrt{\kappa_{2}^2+\tau_{2}^2+1}} & \frac{e^{i \gamma} \left(e^{-i \theta} s_\psi-c_\psi\right)}{\sqrt{b_{0}^2+2}} \\
  \frac{e^{i \gamma} \left(c_\psi-e^{i \theta} s_\psi \tau_{1}\right)}{\sqrt{\kappa_{1}^2+\tau_{1}^2+1}} &\frac{e^{i \gamma} \left(c_\psi+e^{i \theta} s_\psi \tau_{2}\right)}{\sqrt{\kappa_{2}^2+\tau_{2}^2+1}} & \frac{e^{i \gamma} \left(c_\psi+e^{i \theta} s_\psi\right)}{\sqrt{b_{0}^2+2}} \\
\end{array}
\right) \hspace{0.2cm}\mbox{for IH}.
\end{array}%
\right.
\end{equation}
From Eqs. (\ref{KKiNi}) and (\ref{Ulep}), we get:
 \bea s^2_{13}&=&  \left\{
\begin{array}{l}
\fr{\kappa^2_2}{\tau^2_2 + \tau^2_2+1} \hspace{0.3cm}\mbox{for  NH},    \\
\fr{(\tau_{2}+1)^2}{2 \kappa_{2}^2+(\tau_{2}+1)^2} \hspace{0.1cm}\,\mbox{for IH},
\end{array}%
\right. \hs
t^2_{12}=  \left\{
\begin{array}{l}
\fr{\kappa_2^2 (\tau_2-1)^2}{(\tau_2+1)^2 (\kappa_2^2 + \tau_2^2+1)} \hspace{0.2cm}\mbox{for  NH},    \\
\fr{(\tau_{2}-1)^2}{2 \kappa_{2}^2+(\tau_{2}+1)^2} \hspace{0.9cm}\,\mbox{for IH}.
\end{array}%
\right.  \label{thetasq13}\\
  t^2_{23}&=&  \left\{
\begin{array}{l}
\fr{\tau_2^2 c^2_\psi +s^2_\psi-\tau_2 s_{2\psi} c_\theta}{\tau_2^2 s^2_\psi+c^2_\psi +\tau_2 s_{2\psi} c_\theta} \hspace{0.2cm}\mbox{for  NH},    \\
\fr{2}{s_{2\psi} c_\theta+1}-1 \hspace{0.825cm}\,\mbox{for IH}.
\end{array}%
\right. \label{thetasq23}
\eea
 The Jarlskog invariant $J_{CP}$, obtained from Eq. (\ref{Ulep}) and that of the standard parametrization of the lepton mixing
 matrix, leads to the following relation:
 \bea
c_{12} c_{13}^2 c_{23} s_{12} s_{13} s_{23}  s_\delta=\left\{
\begin{array}{l}
-\frac{\kappa_{1} \kappa_{2} (\tau_{1}+\tau_{2}) s_\psi c_\psi s_\theta}{\left(\kappa_{1}^2+\tau_{1}^2+1\right) \left(\kappa_{2}^2+\tau_{2}^2+1\right)} \hspace{0.3cm}\mbox{for  NH},    \\
\hs\hs \frac{b_{0}  \kappa_{2} (1-\tau_{2}) s_\psi c_\psi s_\theta}{\left(b_{0}^2+2\right) \left(\kappa_{2}^2+\tau_{2}^2+1\right)} \hspace{0.75cm}\,\mbox{for IH}.
\end{array}%
\right. \label{Jm}
\eea
The system of equations (\ref{thetasq13})-(\ref{Jm}), with the aid of Eq. (\ref{KKiNi}), yields:
\bea
&&b_0=\left\{
\begin{array}{l}
\frac{\sqrt{2}c_{13}}{\sqrt{s_{13}^2+t_{12}^2}}\hspace{0.25cm}\mbox{for NH},    \\
\sqrt{2} t_{13} \hspace{0.75cm}\mbox{for IH},
\end{array}%
\right. \label{b0expres}\\
&&\kappa_1=\left\{
\begin{array}{l}
\frac{\sqrt{2} c_{13} t_{12} \sqrt{s_{13}^2+t_{12}^2}}{s_{13} \left(t_{12}^2-s_{13} t_{12}+1\right)+t_{12}} \hspace{0.15cm}\mbox{for NH},    \\
\frac{\sqrt{2} c_{13}}{s_{13}+t_{12}} \hspace{2.275cm}\mbox{for IH},
\end{array}%
\right.\,\, \tau_1=\left\{
\begin{array}{l}
1-\frac{2 c^2_{13} t_{12}}{s_{13} \left(t_{12}^2-s_{13} t_{12}+1\right)+t_{12}} \hspace{0.2cm}\mbox{for NH},    \\
1-\frac{2 s_{13}}{s_{13}+t_{12}}\hspace{2.3 cm}\mbox{for IH},
\end{array}%
\right.   \label{ka1tau1expresion}\\
&&\kappa_2=\left\{
\begin{array}{l}
\fr{\sqrt{2} t_{13} \sqrt{s_{13}^2+t_{12}^2}}{s_{13}-t_{12}} \hspace{0.3cm}\mbox{for NH},    \\
\frac{\sqrt{2} c_{13} t_{12}}{1-s_{13} t_{12}}\hspace{1.25cm}\,\mbox{for IH},
\end{array}%
\right. \hspace{1.0 cm} \tau_2=\left\{
\begin{array}{l}
\fr{s_{13}+t_{12}}{s_{13}-t_{12}} \hspace{1.0cm}\mbox{for NH},    \\
\frac{2}{s_{13} t_{12}-1}+1 \hspace{0.2cm}\mbox{for IH},
\end{array}%
\right. \label{K2N2al}\\
&&c_\theta=\left\{
\begin{array}{l}
\frac{[(s_{13} + t_{12})^2 - (s_{13} - t_{12})^2 t_{23}^2] c_\psi^2 +
  [(s_{13} - t_{12})^2 - (s_{13} + t_{12})^2 t_{23}^2] s_\psi^2}{2 c_\psi s_\psi (s^2_{13} - t^2_{12})(1 + t_{23}^2)} \hspace{0.2cm}\mbox{for  NH},\\
\fr{1 - t_{23}^2}{2 c_\psi s_\psi (1+ 2 t_{23}^2)} \hspace{0.25cm}\,\mbox{for IH},
\end{array}%
\right. \label{cthetaexpres} \\
&&s_\delta=
-\frac{ s_{\psi} c_{\psi} s_{\theta}}{s_{23} c_{23} } \hspace{0.3cm}\mbox{for both NH and IH.} \label{sindeltacp}
\eea
Expressions (\ref{b0expres})-(\ref{sindeltacp}) show that $b_0, \kappa_{1,2}$ and $\tau_{1,2}$ depend on two parameters $\theta_{12}$ and $\theta_{13}$ for both NH and IH; $c_\theta$ and $s_\delta$ depend on four parameters $\theta_{12}, \theta_{13}, \theta_{23}$ and $\psi$ for NH while $c_\theta$ and $s_\delta$ depend on three parameters $\theta_{12}, \theta_{13}$ and $\psi$ for IH where $\theta_{12}, \theta_{13}$ and $\theta_{23}$ have been determined with high precision \cite{Salas2021} and $s_\psi$ is a free parameter.

\subsubsection{\label{heavyneutrino} Heavy neutrino mass}
The heavy neutrino mass matrix $\mathrm{M}_R$ in Eq. (\ref{MDR}) has
three eigenvalues and the corresponding eigenvectors as follows
\bea && M_{1} =a_R, \hs M_{2, 3}=\sqrt{b_R^2+d_R^2}\mp c_R, \label{M123}\\
&&\mathbf{U}_N=\left(
\begin{array}{ccc}
 1 & 0 & 0 \\
 0 & i c_N & s_N \\
 0 & -i s_N & c_N \\
\end{array}
\right),\label{UN}\eea
where
\bea
&&c_N\equiv \cos\theta_N=\frac{1}{\sqrt{2}} \sqrt{1+\frac{d_R}{\sqrt{b_R^2+d_R^2}}}, \label{cN}
\label{cN}
\eea
with $b_R$ and $d_R$ are given in Eq. (\ref{abcDR}).
With the aid of Eqs. (\ref{LaSMscale}), (\ref{vevscales}), (\ref{abcDR}) and (\ref{cN}), we can determine the value range of $s_N$ and $c_N$ as follows:
\bea
&& s_{N} \in (0.107,\, 0.280),\hs c_{N} \in (0.959,\, 0.994), \,\,\mathrm{or}, \theta_N\in (6.150, 16.600)^\circ, \label{UN}\eea
i.e., the heavy-heavy neutrino mixing is very small.
\subsubsection{\label{effectiveneutrino} Effective neutrino mass}
We consider firstly the effective neutrino mass ($m_{\mathrm{eff}}$) related to the neutrinoless double beta ($0\nu 2 \beta$) decay which gives the stringent constraint on the mixing
element $\vartheta_{1I} \, (I=1,2,3)$ in the case of the mass of heavy neutrinos $M_{I}$ are about of $\Lambda_\beta=100$ MeV \cite{Benes2005}.
In the SM extension with three right-handed neutrinos, the effective
neutrino mass $m_{\mathrm{eff}}$ is given by \cite{Asakajhep11}
\begin{eqnarray}
&&m_{\mathrm{eff}} = \sum^3_{k=1} U_{1 k}^2 m_k + \sum_{I=1,2,3} M_I \, \vartheta_{e I}^2 \,
  f_\beta(M_I), \label{meegeneral}\end{eqnarray}
where, in Eq. (\ref{meegeneral}), the first term comes from the contribution of active (light) neutrinos while the second term comes from the contribution of heavy (sterile) neutrinos. The function $f_\beta(M_I)$ describes the suppression of the nuclear matrix element from neutrinos with masses heavier than about
100~MeV \cite{Blennowjhep10}.
In the considered model, Eqs. (\ref{LaSMscale}), (\ref{vevscales}), (\ref{abcDR}) and (\ref{M123}) yield the following ranges of heavy neutrino mass $M_I$\bea
&&M_{1} \in (1.25,\, 2.50)10^9\, \mathrm{GeV},\hs M_{2} \in (3.73,\, 5.08)10^9\, \mathrm{GeV}, \hs M_{3} \in (8.92,\, 13.68)10^9\, \mathrm{GeV}, \label{MN}\eea
provided that\footnote{In the case of IH, i.e., $y_1 \in (0.500, 0.750), \, y_2 \in (0.500, 0.750)$ and $y_3 \in (0.022, 0.029)$ we get the ranges of $M_{1,2,3}$: $M_{1} \in (2.50,\, 3.75)10^{10}\, \mathrm{GeV},\, M_{2} \in (3.68,\, 5.57)10^{10}\, \mathrm{GeV}, \, M_{3} \in (3.81,\, 5.70)10^{10}\, \mathrm{GeV}$ which still satisfies the quasi-degenerate condition and $M_I\gg \Lambda_\beta$, i.e., the approximate expression $\left|m_{\mathrm{eff}}\right| \simeq \langle m_{ee}\rangle$ in Eq. (\ref{meeappro}) is satisfied for both NH and IH.}
\bea
&&y_1 \in (2.500, 5.000)10^{-2}, \hs y_2 \in (2.500, 5.000)10^{-2}, \hs y_3 \in (0.115, 0.172). \label{y1y2y3}
\eea
The result in Eq. (\ref{MN}) provides a quasi-degenerate spectrum for heavy neutrino masses. However, 
it also tells us that $M_I$ are at a very high scale with $M_{I}\sim (10^9, 10^{10})$ $\mathrm{GeV}\gg 100\, \mathrm{GeV}=\Lambda_\beta$; thus, $f_\beta
(M_{1,2,3}) \ll 1$ and the contribution of heavy neutrinos to effective neutrino mass is negligible, i.e., the total effective neutrino mass ($m_{\mathrm{eff}}$) is approximately equal to $\langle m_{ee}\rangle$ \cite{Asakajhep11},
\begin{eqnarray}
 \left|m_{\mathrm{eff}}\right| =\langle m_{ee}\rangle= \left|\sum^3_{k=1} U_{1 k}^2 m_k\right|. \label{meeappro}
\end{eqnarray}
Furthermore, Eq. (\ref{m123sq}) provides the relation between neutrino masses as well as their sum and two observed quantities, that is, the two neutrino mass-squared differences $\Delta m^2_{21}$ and $\Delta m^2_{31}$:
\bea
&&\left\{
\begin{array}{l}
\Omega_1=\fr{1}{2}\left(\Delta m^2_{31}+\Delta m^2_{21}\right),\hs \Omega_2=\fr{1}{2}\left(\Delta m^2_{31}-\Delta m^2_{21}\right) \hspace{0.2cm}\mbox{for NH},    \\
\Omega_1=\fr{\Delta m^2_{21}}{2}-\Delta m^2_{31}, \hspace{1.0 cm}  \Omega_2= \fr{\Delta m^2_{21}}{2}\hspace{0.2cm}\mbox{for IH}.
\end{array}%
\right. \label{Om12expr}\\
&&\left\{
\begin{array}{l}
m_1=0, \hs m_2=\sqrt{\Delta m^2_{21}}, \hs m_3=\sqrt{\Delta m^2_{31}} \hspace{0.2cm}\mbox{for NH},    \\
m_1=\sqrt{-\Delta m^2_{31}}, \hs m_2=\sqrt{\Delta m^2_{21}-\Delta m^2_{31}}, \hs m_3=0 \hspace{0.2cm}\mbox{for IH}.
\end{array}%
\right. \label{m1m2m3expresion}\\
&&\sum m_\nu=\left\{
\begin{array}{l}
\sqrt{\Delta m^2_{21}}+\sqrt{\Delta m^2_{31}} \hspace{0.2cm}\mbox{for NH},    \\
\sqrt{\Delta m^2_{21}-\Delta m^2_{31}} +\sqrt{-\Delta m^2_{31}} \hspace{0.2cm}\mbox{for IH}.
\end{array}%
\right. \label{sumexpresion}
\eea
Next, we deal with the effective neutrino masses $\langle m_{ee}\rangle$ and $m_{\beta}$. Combining expressions (\ref{Ulep}), (\ref{b0expres})-(\ref{K2N2al}) and (\ref{m1m2m3expresion}) we obtain the following expressions for $\langle m_{ee}\rangle$ and $m_{\beta}$:
\bea
&&\langle m_{ee}\rangle = \left| \sum^3_{k=1} U_{1 k}^2 m_k \right|
=\left\{
\begin{array}{l}
\sqrt{\Delta m^2_{31}} s_{13}^2+\sqrt{\Delta m^2_{21}} s_{12}^2 c_{13}^2 \hspace{0.2cm} \mbox{for NH, } \\
\left(c_{12}^2 \sqrt{-\Delta m^2_{31}}+s_{12}^2 \sqrt{\Delta m^2_{21}-\Delta m^2_{31}}\right)c_{13}^2 \hspace{0.28cm} \mbox{for IH,}
\end{array}%
\right. \label{meeef}\\
&&m_{\beta } = \sqrt{\sum^3_{k=1} \left|U_{\Revised{1}k}\right|^2 m_k^2}=\left\{
\begin{array}{l}
\sqrt{\Delta m^2_{21} c_{13}^2 s_{12}^2 + \Delta m^2_{31} s_{13}^2} \hspace{0.25cm} \mbox{for NH,} \\
c_{13}\sqrt{\Delta m^2_{21}s_{12}^2 - \Delta m^2_{31}} \hspace{0.75cm} \mbox{for IH.}
\end{array}%
\right.\label{mbef} \eea

\subsection{\label{quark} Quark sector}
From the quark Yukawa terms given in Eq.(\ref{Lquark}) and using the
product rules of the $Q_6$ group, with the help of (\ref{scalarvev}), we obtain the up-and down-quark mass matrices:
\bea M_{u} &=& M_{0u}+\delta M_{u}, \hs M_{d} =M_{0d} +\delta M_{d}, \label{Mud}\eea
where
\bea &&M_{0u} =\left(%
\begin{array}{ccc}
a_u & 0 & 0 \\
0 & b_u+c_u & 0\\
0 & 0   & -b_u+c_u \\
\end{array}%
\right), \hs\hs  \delta M_{u}=\left(%
\begin{array}{ccc}
0 & -g_u & g_u \\
h_u & 0 & -f_u \\
-h_u & f_u    & 0 \\
\end{array}%
\right), \label{MudMu}\\
&&M_{0d} =\left(%
\begin{array}{ccc}
a_d & 0 & 0 \\
0 & b_d+c_d & 0 \\
0 & 0 & -b_d+c_d \\
\end{array}%
\right), \hs\hs  \delta M_{d}=\left(%
\begin{array}{ccc}
0 & -g_d & g_d \\
h_d & 0 & -f_d \\
-h_d & f_d & 0 \\
\end{array}%
\right), \label{MddMd}\eea
with
\bea
&&a_u=\fr{h_{1u}}{\Lambda} v_H v_\rho,\,\, b_u =\fr{h_{2u}}{\Lambda} v_H v_\rho,\,\, c_u=\fr{h_{3u} }{\Lambda} v_H v_\eta, \,\, f_u=\fr{h_{4u}}{\Lambda} v_H v_\phi,\crn
&&g_u=\fr{h_{5u}}{\Lambda} v_H v_\phi,\,\, h_u=\fr{h_{6u}}{\Lambda} v_H v_\phi,\,\, a_d=\fr{h_{1d}}{\Lambda} v_H v_\rho,\,\,\, b_d =\fr{h_{2d} }{\Lambda} v_H v_\rho,\crn
&&c_d=\fr{h_{3d} }{\Lambda} v_H v_\eta,\,\, f_d=\fr{h_{4d}}{\Lambda} v_H v_\phi,\,\, g_d=\fr{h_{5d}}{\Lambda} v_H v_\phi,\,\, h_d=\fr{h_{6d}}{\Lambda} v_H v_\phi. \label{abcgfud}\eea
We found that in Eqs.(\ref{MudMu}) and (\ref{MddMd}), $M_{0u}$ and $M_{0d}$ are due to the
contribution from the scalars $H$, $\rho$ and $\eta$ while $\delta M_{u}$ and $\delta M_{d}$ are due to the contribution from the scalars $H$ and $\phi$. If there is no contribution from $\phi$ then $\delta M_{u,d}$ will vanish and $M_{u}\rightarrow M_{0u},\, M_{d}\rightarrow M_{0d}$, and the quark mixing matrix $V_{CKM}$ being the identity matrix which was ruled out by the recent quark mixing data \cite{PDG2022}. The quark mixing angles are very small, i.e., the quark mixing matrix is close to the unit matrix \cite{PDG2022}. Therefore, we will consider the contribution of $\phi$ as a small perturbation for generating the observed quark mixing angles and terminating the perturbative expansion at the first order.

At the first order of perturbation theory, $\delta M_{u}$ and $\delta M_{d}$ have no contribution to the eigenvalues, however, they make contributions and thus change the corresponding eigenvectors. The quark masses and mixing matrices are given by:
\bea
&&m_{u} =a_{u},\hs m_{c}=b_{u} + c_{u},\hs m_t=-b_{u}+c_{u}, \crn
&&m_{d}= a_{d},\hs m_{s}=b_{d} + c_{d},\hs m_b=-b_{d} + c_{d}, \label{quarkmasses}\\
&&U_{u, d}=\left(
\begin{array}{ccc}
 1 & \frac{g_{u, d}}{x_{u, d}} & -\frac{g_{u, d}}{y_{u, d}} \\
 \frac{h_{u, d}}{x_{u, d}} & 1 & \frac{f_{u, d}}{2 b_{u, d}} \\
-\frac{h_{u, d}}{y_{u, d}} & \frac{f_{u, d}}{2 b_{u, d}} & 1 \\
\end{array}
\right),\label{UudLR}
 \eea
 where
$a_{u,d}, b_{u,d}$, $c_{u,d}$, $f_{u,d}$, $g_{u,d}$ and $h_{u,d}$ are defined in Eq.(\ref{abcgfud}), and
\bea
&&x_{u} = a_{u} - b_{u} - c_{u}, \hs
y_{u} = a_{u} + b_{u} - c_{u}. \label{albetaud}
\eea
Eq. (\ref{quarkmasses}) requires $a_{u,d}, b_{u,d}$ and $c_{u,d}$ to be real numbers. This can be satisfied by considering the corresponding Yukawa-like couplings are real. The $U_\mathrm{CKM}$ 
matrix is defined as, $U_\mathrm{CKM}= U_{uL} U^{\dagger}_{dL}$, which owns the following entries:
\bea
&&U_\mathrm{CKM}^{11}=  1+g^{*}_d g_{u} \left(\fr{1}{x_{d} x_{u}}+\fr{1}{y_{d} y_{u}}\right),\hs\,
U_\mathrm{CKM}^{12}= \fr{h^*_{d}}{x_{d}}+\fr{g_{u}}{x_{u}}-\fr{f^*_{d} g_{u}}{2 b_{d} y_{u}},\crn
&&U_\mathrm{CKM}^{13}=-\fr{h^*_{d}}{y_{d}}-\fr{g_{u}}{y_{u}}+\fr{ f^*_{d} g_{u}}{2 b_{d} x_{u}},\hspace{1.35 cm}
U_\mathrm{CKM}^{21}= \fr{h_{u}}{x_{u}}+\fr{g^*_{d}}{x_{d}}-\fr{f_{u} g^*_{d}}{2 b_{u} y_{d}}, \crn
&&U_\mathrm{CKM}^{22}= 1+\fr{f^*_{d} f_{u}}{4 b_{d} b_{u}}+\fr{h^*_{d} h_{u}}{x_{d} x_{u}},\hspace{1.6 cm}
 U_\mathrm{CKM}^{23}=\fr{1}{2} \left(\fr{f^*_{d}}{b_{d}}+\fr{f_{u}}{b_{u}}\right)-\fr{h^*_{d} h_{u}}{x_{u} y_{d}},\crn
&&U_\mathrm{CKM}^{31}=-\fr{h_{u}}{y_{u}}-\fr{g^*_{d}}{y_{d}}+\fr{f_{u} g^*_{d} }{2 b_{u} x_{d}}, \hspace{1.35 cm}
U_\mathrm{CKM}^{32} =\fr{1}{2} \left(\fr{f_{u}}{b_{u}}+\fr{f^*_{d}}{b_{d}}\right)-\fr{h^*_{d} h_{u}}{x_{d} y_{u}}, \crn
&&U_\mathrm{CKM}^{33}=1+\fr{f^*_{d} f_{u}}{4 b_{d} b_{u}}+\fr{h^*_{d} h_{u}}{y_{d} y_{u}}. \label{Uckmv}\eea
As we will show in Section \ref{NR}, the experimental data on quark mass and mixing angles \cite{PDG2022} show that the model's results on Eqs. (\ref{quarkmasses}) and (\ref{Uckmv}) are consistent since the quark Yukawa-like couplings $a_{u,d}, b_{u,d}$, $c_{u,d}$, $f_{u,d}$, $g_{u,d}$ and $h_{u,d}$ can be fitted with the observed data.
\subsection{\label{NR} Numerical analysis and discussion}
\emph{$\bullet$ For the charged lepton sector}, expressions
 (\ref{h1h2})-(\ref{Delh}) tell us that $|h_{1}|$ depends on $\La, m_e, v_\eta$ and $v_S$; $|h_{2}|$ depends on $\Lambda, m_\mu, m_\tau, v_H, v_\eta, \psi$ and $\theta$; $|h_{3}|$ depends on $\Lambda, m_\mu, m_\tau, v_H, v_\rho, \psi$ and $\theta$; and $|h_{4}|$ depends on $\Lambda, m_\mu, m_\tau, v_H, v_\phi, \psi$ and $\theta$ where $\theta$ is expressed as a function of four parameters $\theta_{12}, \theta_{13}, \theta_{23}$ and $\psi$ as given
in  Eq. (\ref{cthetaexpres}). For the given VEV scales of scalar fields and the cut-off scale in Eqs (\ref{LaSMscale})-(\ref{vevscales}) and the charged lepton masses $m_{e,\mu,\tau}$ \cite{PDG2022}, there exist possible range of the model parameter $\psi$ and $\theta_{12}, \theta_{13}, \theta_{23}$ such that the Yukawa couplings in the charged lepton sector, $h_{1,2,3,4}$, differ by only one order of magnitude, i.e., the charged lepton mass hierarchy is satisfied. Firstly, considering the case of $v_H>v_S$, the relation $v_H^2+v_S^2 =246^2 \, \mathrm{GeV}^2$ yields $v_H\in (174, 246)\, \mathrm{GeV}$. As a consequence, we obtain $h_1\in (5.88\times 10^{-5}, 4.61\times 10^{-2})$, $h_2\in (7.15, 7.65)10^{-2}$, $h_3\in (4.16, 9.47)10^{-2}$, $h_4\in (2.10, 7.93)10^{-2}$, i.e., $h_{2,3}$ and $h_4$ are in the same scale $\sim 10^{-2}$ while $h_1$ has much smaller values than $h_{2,3,4}$ when it decreases toward $174$ GeV. $h_1$ can be of the same order as $h_{2,3,4}$ when $v_H$ increases to very close to $246$ GeV. Namely, $h_1$ can reach a range of $(0.421,  4.61)10^{-2}$ in the case of $v_H\in (245.988,\, 246.0)$ GeV. Therefore, in this study, we consider the case of $v_S \simeq 2.46\, \mathrm{GeV}, v_H\simeq 245.988\, \mathrm{GeV}$, i.e., $v_S$ is less than $v_H$ about two order of magnitude as in Eq. (\ref{vSvH}). Next, using the observed values of the charged lepton masses \cite{PDG2022}, $m_e=0.51099 \,\mathrm{MeV},  m_\mu = 105.65837\,\mathrm{MeV}, m_\tau = 1776.86 \,\mathrm{MeV}$, the values of $\Lambda$ and the VEV of scalar fields in Eqs. (\ref{LaSMscale})-(\ref{vevscales}), with the help of Eqs. (\ref{h1h2})-(\ref{Delh}) and (\ref{cthetaexpres}), we get $|h_1|=4.15\times 10^{-3}$, and $|h_{2}|, |h_{3}|$ and $|h_{4}|$ depend on four parameters $\theta_{12}, \theta_{13}, \theta_{23}$ and $\psi$ where $s^2_{12}, s^2_{13}$ and $s^2_{23}$ have been measured with high precision \cite{Salas2021}.
In the consided model, we find that the possible domains of $\theta_{12}, \theta_{13}, \theta_{23}$ which belong to 3$\sigma$ experimentally allowed ranges taken from Ref. \cite{Salas2021},
\bea
&&s^2_{13} \in (2.0\times 10^{-2},\, 2.2\times 10^{-2}), \hs
s^2_{23} \in (0.584,\, 0.610), \hs s^2_{12}\in (0.344,\, 0.369). \label{thetaijjdomain}\eea
Besides, Eqs. (\ref{cthetaexpres}) and (\ref{sindeltacp}) provide the relations between $\delta_{CP}, \psi, \theta$ and lepton mixing angles. However, the experimental constraint on $\delta_{CP}$ still has a quite large error \cite{Salas2021}. Therefore, we do not consider $\delta_{CP}$ as an observed parameter but consider $\psi$ as an input parameter to determine the range of values of $\delta_{CP}$. In order to make the Yukawa couplings $|h_{2}|, |h_{3}|$ and $|h_{4}|$ 
in the same scale of magnitude and differ by only one order of magnitude compared to $|h_{1}|$, the parameter $\psi$ is adjusted as follows:
\bea
&& s_\psi \in (0.150,\, 0.440),\,\,\mathrm{i.e.},\,\, \psi\in (8.630,\, 26.100)^{\circ}.  \label{spsidomain}
\eea
It is noted that in the case of $s_\psi<0.15$ or $s_\psi>0.44$, $|h_4|$ being complex numbers; thus, they are rule out.
With the aid of Eqs. (\ref{LaSMscale})-(\ref{vevscales}), (\ref{h1h2})-(\ref{Delh}), (\ref{cthetaexpres}), (\ref{thetaijjdomain}) and (\ref{spsidomain}) we get the following ranges for $|h_{2}|, |h_{3}|$ and $|h_{4}|$:
\bea
&&|h_2|\in (7.150,\, 7.650)10^{-2}, \hs |h_3|\in (4.160, 6.690)10^{-2}, \hs |h_4|\in (2.100, 5.610)10^{-2}. \label{h2h3h4ranges}
\eea
\emph{$\bullet$ For neutrino sector}, expressions (\ref{b0expres})-(\ref{sindeltacp}) and (\ref{m1m2m3expresion})-(\ref{mbef}) show that
$b_{0}$ depends on $\theta_{12}$ and $\theta_{13}$ for NH and 
depends on $\theta_{13}$ for IH; $\kappa_{1,2}$ and $\tau_{1,2}$ depend on $\theta_{12}$ and $\theta_{13}$ for both NH and IH; $c_{\theta}$ and $s_{\delta}$ depend on $\theta_{12}$, $\theta_{13}, \theta_{23}$, $\psi$ for NH and depend on $\theta_{23}, \psi$ for IH. On the other hand, $m_{2}$ depends on $\Delta  m^2_{21}$, $m_{3}$ depends on $\Delta m^2_{31}$ for NH while $m_{1}$ depends on $\Delta m^2_{31}$, $m_{2}$ depends on $\Delta m^2_{21}$ and $\Delta m^2_{31}$ for IH, and $\sum m_{\nu}$  depends on $\Delta  m^2_{21}$ and $\Delta m^2_{31}$ for both NH and IH; $\langle m_{ee}\rangle$ and $m_\beta$ depend on $\Delta  m^2_{21}, \Delta m^2_{31}, \theta_{12}$ and $\theta_{13}$ for both NH and IH. In order to determine the possible ranges of the model parameters $\theta, \psi, b_0, \kappa_{1,2}$, $\tau_{1,2}$ and get possible values for the observable parameters $s_\delta$, $m_{1,2,3}$, $\sum m_\nu$, $\langle m_{ee} \rangle$ and $m_{\beta}$, we use the neutrino mixing angles $s^2_{12}, s^2_{13}$ and $s^2_{23}$ given in Eq. (\ref{thetaijjdomain}) and the neutrino mass-squared differences $\Delta m^2_{21}$, $\Delta m^2_{31}$  whose experimental values within 3 $\sigma$ range are taken from Ref. \cite{Salas2021} with
\bea
&&\left\{
\begin{array}{l}
 \Delta m^2_{21}\in  (69.40,\,  81.40)\,\mathrm{meV}^2 \hspace{0.2cm}\mbox{for  both NH and IH},  \\
\Delta m^2_{31}\in  (2.47,\,  2.63)10^3 \, \mathrm{meV}^2\hspace{0.15cm}\mbox{for NH},\hs \Delta m^2_{31}\in  (-2.53,\,  -2.37)10^3 \, \mathrm{meV}^2\hspace{0.15cm}\mbox{for IH}. \label{squaredmass}
\end{array}%
\right.\eea
With the aid of Eqs. (\ref{b0expres})-(\ref{sindeltacp}) and (\ref{thetaijjdomain})-(\ref{spsidomain}), we find the possible ranges of $b_0, \kappa_{1,2}, \tau_{1,2}$, $c_\theta$ and $s_\delta$ as follows
\bea
&&b_0\in  \left\{
\begin{array}{l}
(1.795, \, 1.897)\, \hspace{0.20cm}\mbox{for  NH},  \\
(0.202, 0.212) \,\hspace{0.25cm}\mbox{for IH},
\end{array}%
\right. \crn
&& \kappa_{1}\in \left\{
\begin{array}{l}
(0.801, 0.855) \hspace{0.2cm}\mbox{for  NH},  \\
(1.532, 1.617)\hspace{0.25cm}\mbox{for IH},
\end{array}%
\right. \hspace{0.55 cm} \kappa_{2}\in \left\{
\begin{array}{l}
(-0.272, -0.252) \hspace{0.75cm}\mbox{for  NH},  \\
(-1.206, -1.129)\hspace{0.2cm}\mbox{for IH},
\end{array}%
\right. \\
&&\tau_{1}\in \left\{
\begin{array}{l}
(-0.540, -0.516) \hspace{0.15cm}\mbox{for  NH},  \\
(0.660,\, 0.688)\hspace{0.75cm}\mbox{for IH},
\end{array}%
\right. \hspace{0.1 cm} \tau_{2}\in \left\{
\begin{array}{l}
(-1.515,-1.454) \hspace{0.25cm}\mbox{for  NH},  \\
(-1.256, -1.228)\hspace{0.25cm}\mbox{for IH}.
\end{array}%
\right.\label{KKN12values}\\
&&c_{\theta}\in \left\{
\begin{array}{l}
(-0.761, 1.139\times 10^{-3}) \hspace{0.25cm}\mbox{for  NH},  \\
(-0.742, -0.213)\hspace{0.25cm}\mbox{for IH},
\end{array}%
\right. \mbox{i.e.},\, \theta\in \left\{
\begin{array}{l}
(89.93, 139.50)^\circ \hspace{0.25cm}\mbox{for  NH},  \\
(102.00, 138.00)^\circ \hspace{0.15cm}\mbox{for IH},
\end{array}%
\right. \label{c1range}\\
&&s_\delta\in \left\{
\begin{array}{l}
(-0.810, -0.195)\hspace{0.15cm}\mbox{for  NH},  \\
(-0.783, -0.204)\hspace{0.15cm}\mbox{for IH},
\end{array}%
\right. \, \mbox{i.e., \,} \delta_{CP}\in \left\{
\begin{array}{l}
(305.90, 348.70)^\circ \hspace{0.2cm}\mbox{for  NH},  \\
(308.00, 348.00)^\circ \hspace{0.2cm}\mbox{for IH}.
\end{array}%
\right.\label{sdrange}
\eea
The obtained ranges of the Dirac CP violation phase in Eq. (\ref{sdrange}) belong to the $3\sigma$ experimentally allowed range of the best-fit vales taken from Ref.\cite{Salas2021} for both NH and IH.

Now we come back to the neutrino mass issue. Eq. (\ref{m1m2m3expresion}) shows that neutrino masses
$m_{2,3}$ for NH ($m_{1,2}$ for IH) depend on two experimental parameters $\Delta m^2_{21}$ and $\Delta m^2_{31}$ which have been measured with high accuracy.
Within 3$\sigma$ range of $\Delta m^2_{21}$ and $\Delta m^2_{31}$ taken from Ref. \cite{Salas2021} as given in Eq. (\ref{squaredmass}), we get the allowed regions for neutrino masses and their sum as follows:
\bea
&&\left\{
\begin{array}{l}
m_1=0,\hs m_2\in  (8.33, 9.02)\, \mathrm{meV},\hs m_3= (49.70, 51.28) \, \mathrm{meV} \hspace{0.15cm}\mbox{for  NH},    \\
m_1\in (48.68, 53.30)\, \mathrm{meV},\,  m_2= (49.39,51.10)\, \mathrm{meV}, \hs m_3=0 \hspace{0.25cm}\mbox{for  IH},
\end{array}%
\right. \label{mirange}\\
&&\sum m_{\nu} \in \left\{
\begin{array}{l}
(58.03, 60.51)\, \mathrm{meV} \hspace{0.2cm}\mbox{for  NH},    \\
(98.07, 101.40)\, \mathrm{meV} \hspace{0.2cm}\mbox{for  IH}.
\end{array}%
\right. \label{sumrange}
\eea
There are recently different constraints for the sum of neutrino mass such as the updated bounds in Ref.\cite{RoyChoudhury} with $\sum m_{\nu}< 0.15$ eV for NH and $\sum m_{\nu}< 0.17$ eV for IH,
$\sum m_{\nu} < 0.14 $ eV \cite{Tanseri22}, $\sum m_{\nu} < 0.139$ eV (NH) and $\sum m_{\nu} < 0.174$ eV (IH) \cite{Stockerprd21}, $\sum m_{\nu} < 0.099$ eV  (Planck+BAO+RSD+SNe)\,\cite{Alamprd21}, $\sum m_{\nu} < 0.42$ eV \cite{GuillermoJHEP21}, $\sum m_{\nu} < 0.113$ eV for NH and $\sum m_{\nu} < 0.145$ eV for IH \cite{Adame2024arxiv}\footnote{The \mbox{DESI BAO+CMB} gives an upper 
constraint of $\sum m_\nu < 0.072$ eV in Ref. \cite{Adame2024arxiv} obtained  by choosing the minimal mass $\sum m_{\nu}>0$ eV is chosen. However, the experimental data in Ref. \cite{Salas2021} implies a minimal mass $\sum m_{\nu}>0.058$ eV for NH and $\sum m_{\nu}>0.098$ eV for IH. Therefore, the corresponding total masses are $\sum m_{\nu} < 0.113$ eV for NH and $\sum m_{\nu} < 0.145$ eV for IH \cite{Adame2024arxiv}.}. The obtained results on the sum of neutrino mass in Eq. (\ref{sumrange}) are
consistent with all the limits taken from Refs.\cite{RoyChoudhury,Tanseri22, Alamprd21,Stockerprd21, GuillermoJHEP21, Adame2024arxiv} excepting the combination Planck+BAO+Ly-$\alpha$ with $\sum m_{\nu} < 0.089$ eV \cite{Jimenezjcap22} for IH.

Expressions (\ref{meeef})-(\ref{mbef}) show that $\langle m_{ee}\rangle$ and $m_{\beta }$ depend on four parameters $\theta_{12}, \theta_{13}$, $\Delta m^2_{21}$ and $\Delta m^2_{31}$ where $\theta_{12}, \theta_{13}$ and $\Delta m^2_{21}$ have been determined with high precision and $\Delta m^2_{31}$ depends on the neutrino mass hierarchy \cite{Salas2021}. With the aid of Eqs. (\ref{thetaijjdomain}) and (\ref{squaredmass}), we find the following ranges  for the effective neutrino masses,
\bea
&&\langle m_{ee}\rangle \in  \left\{
\begin{array}{l}
(3.80, 4.38) \, \mbox{meV} \hspace{0.25cm} \mbox{for NH,} \\
(47.85, 49.58) \, \mbox{meV} \hspace{0.25cm} \mbox{for  IH,}
\end{array}%
\right.    \label{meerange}\\
&& m_{\beta} \in\left\{
\begin{array}{l}
(8.53, 9.34)\, \mbox{meV} \hspace{0.25cm} \mbox{for NH,} \\
(48.39, 50.09)\, \mbox{meV} \hspace{0.25cm} \mbox{for  IH}.
\end{array}%
\right.   \label{mbrange}
\eea
These predictive ranges are below the upper limits for effective neutrino mass related to $0\nu \beta \beta$ decay from KamLAND-Zen \cite{KamLAND16} $\langle m_{ee} \rangle <61-165\, \mathrm{meV}$, GERDA \cite{GERDA19} $\langle m_{ee} \rangle < 104-228\, \mathrm{meV}$, and CUORE \cite{CUORE20} $\langle m_{ee} \rangle < 75-350 \,\mathrm{meV}$ as well as the upper limits for effective neutrino mass related to $\beta$-decay from \cite{PDG2022} with $8.5 \,\mathrm{meV} < m_{\beta} < 1.1\, \mathrm{eV}$ for NH and $48 \, \mathrm{meV} < m_{\beta} < 1.1\, \mathrm{eV}$ for IH, $8.90\, \mathrm{eV} <m_{\beta} < 12.60\, \mathrm{eV}$ \cite{mbet3constraint}, and $m_{\beta} < 0.8\, \mathrm{eV}$ \cite{Aker22n}. The meV limit of $\langle m_{ee}\rangle$ can be reached by the future experiments \cite{JCao2020mev}.

Next, we determine the order of magnitude of the Yukawa-like couplings in neutrino sector. For this purpose, we assume
\bea
|x_2|= r |x_1|, \hs |y_3|=p |y_2|, \label{xyrelation}\eea
with $r$ and $p$ are the scaling factors.
By combining Eqs. (\ref{LaSMscale})-(\ref{vevscales}), (\ref{abcDR}),(\ref{ab}),(\ref{XYZ})-
(\ref{m123sq}), (\ref{Om12}), (\ref{Unu}) and (\ref{Om12expr})-(\ref{m1m2m3expresion}),
we get
\bea
&&\left\{
\begin{array}{l}
r=3.240,\hs\hs p=8.220 \hspace{0.2cm} \mbox{for NH,} \\
r=11.450, \hs p=0.230 \hspace{0.25cm} \mbox{for IH.}
\end{array}%
\right.
\eea
The possible values of $|x_{1}|$, $|x_{2}|$, $|y_{2}|$ and $|y_{3}|$ satisfying the neutrino mass hierarchy are
\bea
&&|x_{1}| \in  \left\{
\begin{array}{l}
(3.530, \,4.990)10^{-2} \hspace{0.25cm} \mbox{for NH,} \\
(1.250, 1.530)10^{-2} \hspace{0.275cm} \mbox{for IH,}
\end{array}%
\right.    \hs |x_{2}| \in \left\{
\begin{array}{l}
(0.114, \,0.162)\, \mbox{meV} \hspace{0.25cm} \mbox{for NH,} \\
(0.143, \,0.176)\, \mbox{meV} \hspace{0.25cm} \mbox{for IH},
\end{array}%
\right. \label{x1x2range}\\
&&|y_{2}| \in \left\{
\begin{array}{l}
(2.500,\, 5.000)10^{-2} \hspace{0.15cm} \mbox{for NH,} \\
(0.500, 0.750) \hspace{1.0cm} \mbox{for IH},
\end{array}%
\right. \hs\hs  |y_{3}| \in \left\{
\begin{array}{l}
(0.115,\, 0.172) \hspace{0.25cm} \mbox{for NH,} \\
(0.022, 0.029) \hspace{0.35cm} \mbox{for IH},
\end{array}%
\right. \label{y2y3range}
\eea
which differ by only one order of magnitude.

\emph{$\bullet$ For the quark sector}, comparing the quark masses and the elements of $U_{\mathrm{CKM}}$ with their corresponding best-fit points \footnote{The best-fit points in Table \ref{quarkpara} correspond to the Wolfenstain parameters\cite{PDG2022}: $\lambda=0.2250,\hs A=0.826, \hs \bar{\rho}=0.159$ and $\bar{\eta}=0.348$ which correspond to the mixing angles $\sin \theta^{q}_{12} = 0.22500, \hs \sin \theta^{q}_{13}= 0.00369, \hs \sin \theta^{q}_{23}=0.04182$ and $\delta^{q}_{CP}=1.444$ \cite{PDG2022}.} given in Refs.\cite{PDG2022}, we obtain a prediction for the quark mixing matrix and the model's parameters in the quark sector as shown in Table \ref{quarkpara} and Eqs. (\ref{modelpara})-(\ref{modelpara1}), respectively.
\begin{table}[tbh]
\caption{\label{quarkpara}The best-fit points for quark parameters taken from Ref.\cite{PDG2022} and the model prediction.}
\vspace{-0.25cm}
\begin{center}
\begin{tabular}{|c|c|c|c|}
\hline
Observable & Best-fit point \cite{PDG2022} & The model prediction& Percent error $(\%)$ \\ \hline\hline
$m_u [\mathrm{MeV}]$  &  \quad $2.16$ &$2.16$& $0$ \\ \hline
$m_c [\mathrm{GeV}]$ & \quad $1.27$ &$1.27$& $0$ \\ \hline
$m_t [\mathrm{GeV}]$&  \quad $172.69$ &$172.69$& $0$ \\ \hline
$m_d [\mathrm{MeV}]$&  \quad $4.67$ &$4.67$& $0$ \\ \hline
$m_s [\mathrm{MeV}]$&  \quad $93.4$ &$93.4$& $0$ \\ \hline
$m_b [\mathrm{GeV}]$ &  \quad $4.18$ &$4.18$& $0$ \\ \hline\hline
$U_{\mathrm{CKM}}^{11}$ &  \quad $0.974352$ &  \quad $0.974352$& $0$\\ \hline
$U_{\mathrm{CKM}}^{12}$ &  \quad $0.224998$ &  \quad $0.224998$& $0$\\ \hline
$U_{\mathrm{CKM}}^{13}$ &  \quad $0.0015275 - 0.003359 i$ &   $-0.0014739 - 0.0033507 i$& $0.804$ \\ \hline
$U_{\mathrm{CKM}}^{21}$ &  \quad $-0.224865$&  $-0.224865$& $0$ \\ \hline
$U_{\mathrm{CKM}}^{22}$ &  \quad $0.973492$&   $0.973492$ & $0$\\ \hline
$U_{\mathrm{CKM}}^{23}$ &  \quad $0.0418197$ &$-0.0407346$& $2.66$\\ \hline
$U_{\mathrm{CKM}}^{31}$ &  \quad $0.0079225 - 0.00327 i$ &$0.0095908 + 0.0080436 i$& $31.50$\\ \hline
$U_{\mathrm{CKM}}^{32}$ &  \quad $-0.041091 $&$-0.041091$ &$0$\\ \hline
$U_{\mathrm{CKM}}^{33}$ &  \quad $0.999118$ &$0.999118$& $0$ \\ \hline
\end{tabular}%
\end{center}
\vspace{-0.5cm}
\end{table}
\bea
&&a_{u}=2.16\times 10^{-3} \,\mathrm{GeV}, \hs
b_{u}=-85.71\, \mathrm{GeV}, \hs c_u=86.98\, \mathrm{GeV}, \crn
&&g_u=(-0.143+0.145 i)\,\mathrm{GeV}, \, f_u=9.727\, \mathrm{GeV}, \, h_u=(0.142-0.145 i)\, \mathrm{GeV}, \crn
&&a_{d}=4.67\times 10^{-3} \,\mathrm{GeV}, \hs b_{d}=-2.04\, \mathrm{GeV}, \hs c_d=2.14\, \mathrm{GeV}, \crn
&&g_d=(9.972-10.10 i) 10^{-3}\, \mathrm{GeV}, \hs f_d=-6.32\times 10^{-2}\, \mathrm{GeV}, \crn
&&h_d=(-9.974+10.12i) 10^{-3}\, \mathrm{GeV}. \label{modelpara}
\eea
Next, comparing Eqs. (\ref{abcgfud}) and (\ref{modelpara}) with the aid of Eqs. (\ref{LaSMscale})-(\ref{vevscales}), one obtains:
\bea
&&|h_{1u}|\simeq 10^{-4}, \hs |h_{2u}|\sim |h_{3u}|\simeq 7.0, \hs\,\, |h_{4u}|\simeq 1.0,  \hs\,\, |h_{5u}|\sim |h_{6u}|\simeq 10^{-2},\crn
&& |h_{1d}|\simeq 10^{-4}, \hs |h_{2d}|\sim |h_{3d}|=10^{-1},\, |h_{4d}|\simeq 10^{-2}, \,
|h_{5d}|\sim  |h_{6d}|=10^{-3}. \label{modelpara1}
\eea
The results in Table \ref{quarkpara} imply that all the obtained values of the CKM matrix are in agreement with the global fit taken from Ref. \cite{PDG2022}, except for the "31" element. However, the obtained value of "31" element with $|U_{\mathrm{CKM}}^{31}|=0.0125$ is consistent with the 90$\%$ CL limit taken from Ref. \cite{Renk2002proceeding} by using the eight tree-level constraints with $|U_{\mathrm{CKM}}^{31}| \in (0.004, 0.014)$, and $|U_{\mathrm{CKM}}^{31}|\in (0.000, 0.038)$ \, (marginal) and $|U_{\mathrm{CKM}}^{31}|\in (0.000, 0.023)$ \, (individual) \cite{Cruz24arxiv, Clerbauxjhep2019}.
\section{\label{anomal}Muon anomalous magnetic moment}

The most recent experimental result for $(g-2)_\mu$~\cite{Muong-2:2023cdq} reported by the Muon $g-2$ Collaboration, improves its previous result by more than a factor of two \cite{Abimug2}, potentially showing deviations from the prediction of the Standard Model~\cite{Aoyamag2}. The experimentally measured value the anomalous magnetic moment of the muon is given by \cite{Muong-2:2023cdq}\footnote{The SM value of the muon anomalous magnetic moment yielding the $5.1\sigma$ discrepancy is based on the data-driven determination of the hadronic vacuum polarization contributions to $a_{\mu}^{\text{SM}}$. Nevertheless, recent lattice QCD calculations of the same quantity consistently point toward a significantly higher value \cite{FermilabLattice:2019ugu,Borsanyi:2020mff,Lehner:2020crt,Aubin:2019usy}, which would result in reduction of the muon $g-2$ anomaly down to $1.5\sigma$ \cite{Wittig:2023pcl}.}:
\be
\Delta a_{\mu} = a_{\mu}^{\text{exp}}-a_{\mu}^{\text{SM}} = (2.49\pm 0.48) \times  10^{-9}.\label{eq:gm2}
\ee
In this section we will analyze the implications of our model in the muon anomalous magnetic moment to one loop and two loop.

\begin{center}
\begin{figure}[h]
\vspace{-3.25cm}
\hspace{-1.0cm}\includegraphics[width=0.8\textwidth]{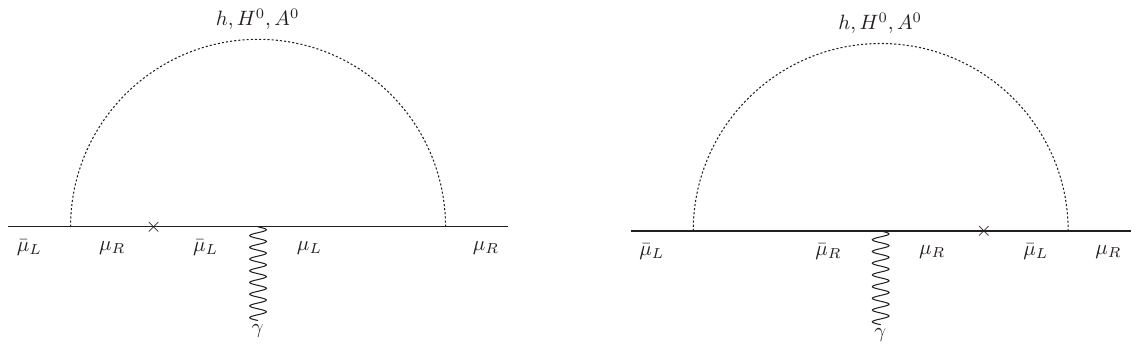}\hspace{-1.0 cm}
\vspace{-11.25cm}
\caption{One loop Feynman diagrams contributing to the muon anomalous magnetic moment.}
\label{fig:delta_mau2}
\end{figure}
\vspace{0.15 cm}
\end{center}

In one loop, the muon anomalous magnetic moment receives contributions arising from vertex diagrams involving the exchange of neutral scalars and charged leptons running in the internal lines of the loop. Then, in our model, the contributions to the muon anomalous magnetic moment take the form \cite{Diaz:2002uk}:
\bea
\Delta a_{\mu} &\simeq & \fr{\left[(y_{h_{\overline{\mu}_{R}\mu_L}})^2-(y_{h_{\overline{\mu}_{R}\mu_L}}^{\text{SM}})^2\right]m_{\mu}^2}{8\pi^2}I_H^{(\mu)}(m_{\mu},m_h)\notag \\
&+&\fr{m_{\mu}^2}{8\pi^2}\left[y_{H_{\overline{\mu}_{R}\mu_L}^0}^2I_H^{(\mu)}(m_{\mu},m_{H^0})+y_{A_{\overline{\mu}_{R}\mu_L}^0}^2I_A^{(\mu)}(m_{\mu},m_{A^0}) \right],\label{eq:m_au}
\eea
where the loop function $I_{H(A)}^{(\mu)}(m_f,m_{H(A)})$ has the form:
\be
I_{H(A)}^{(\mu)}(m_f,m_{H(A)}) = \int_0^1\fr{x^2\left(1-x\pm \fr{m_f}{m_{\mu}}\right)}{m_{\mu}^2x^2+(m_f^2-m_{\mu}^2)x+m_{H(A)}^2(1-x)}dx\Revised{,}
\ee
with the sign "$+$" is for CP-even scalars and the sign "$-$" is for CP-odd scalars.
\begin{center}
\begin{figure}[h]
\vspace{0.75cm}
\vspace{-2.65cm}
\hspace{-1.0cm}\includegraphics[width=0.7\textwidth]{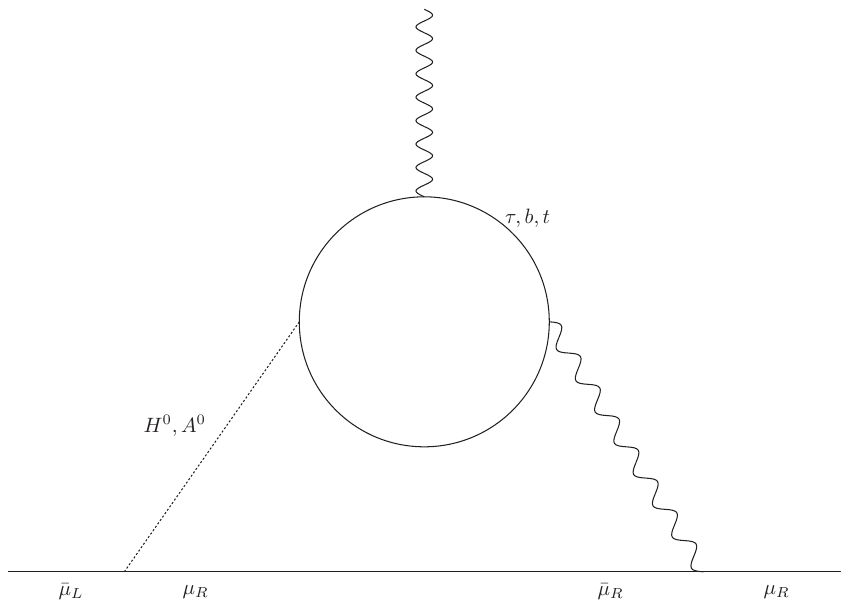}\hspace{-1.0 cm}
\vspace{-8.0cm}
\caption{Two loop Feynman Bar-Zee diagram contributing to the muon anomalous magnetic moment.}
\label{fig:delta_mau2}
\end{figure}
\vspace{-0.75 cm}
\end{center}
\begin{center}
\begin{figure}[h]
\vspace{-0.5 cm}
\hspace{-1.0 cm}
\hspace{-1.0 cm}\includegraphics[width=0.6\textwidth]{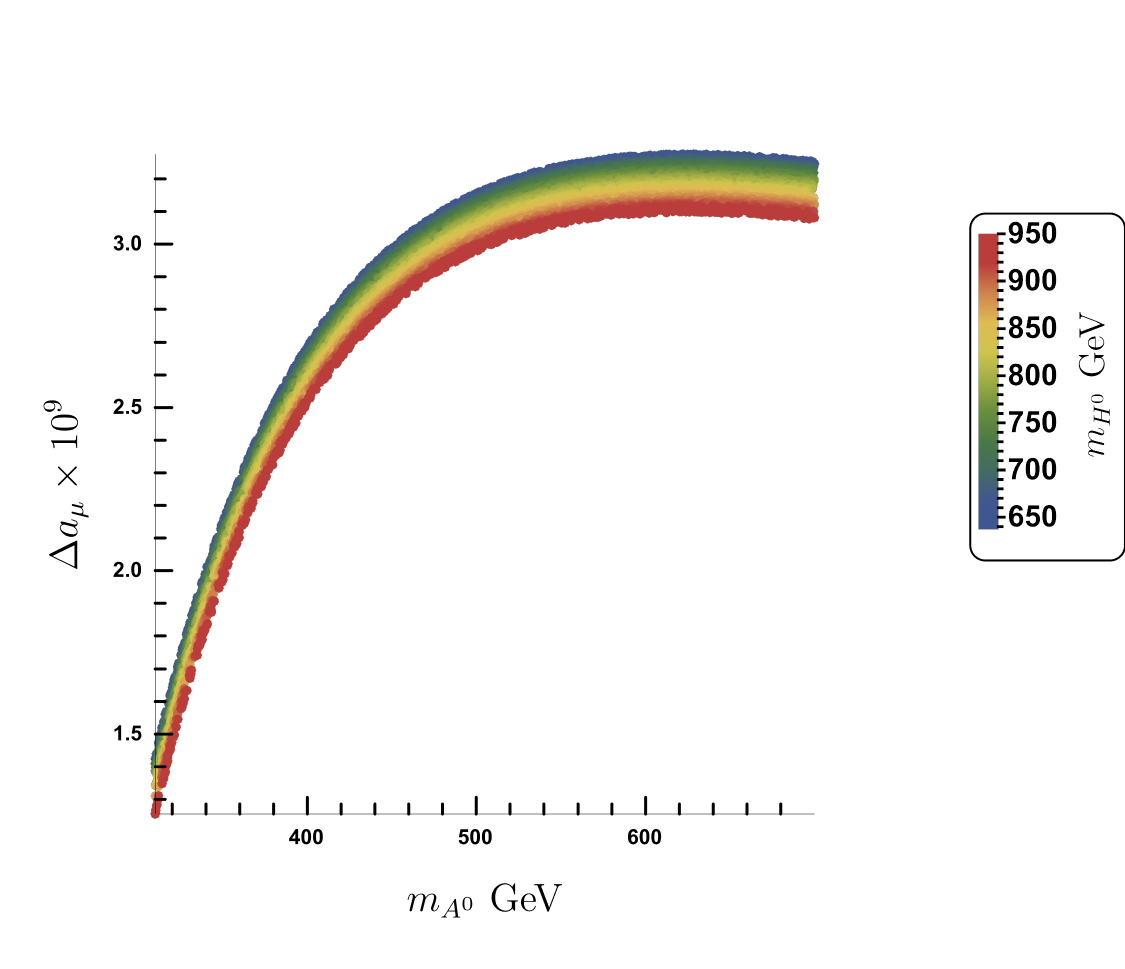}\hspace{-1.0 cm}
\vspace{-0.45 cm}
\caption{Correlation plot between the mass of the CP-odd scalar $A^0$ and the deviation of the anomalous magnetic moment of the muon $\Delta a_{\mu}$, for different values of the CP-even scalar $m_{H^0}$.}
\label{fig:delta_mau2}
\end{figure}

\end{center}
\begin{figure}[h]
\vspace{0.65 cm}
\centering
\includegraphics[width=0.75\textwidth]{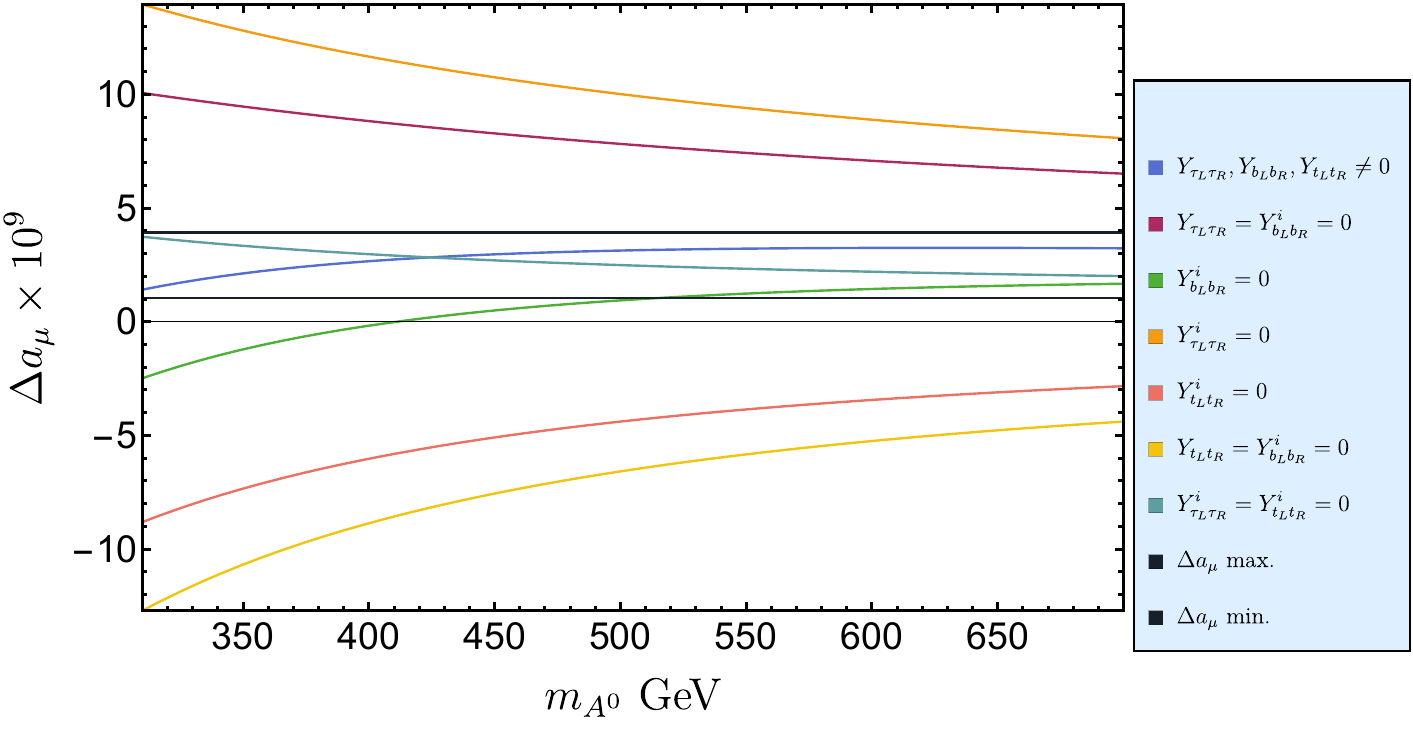}
\caption{Deviation of the anomalous magnetic moment of the muon $\Delta a_{\mu}$ as a function of the mass of the CP-odd scalar $A^0$, considering the mass of the CP-even scalar $m_{H^0}=700\ \text{GeV}$. The colored curves represent different benchmarks for the same mass of the CP-even scalar, where the right box indicates the couplings that have been considered zero in each curve, while the black horizontal lines represent the maximum and minimum experimental values for $\Delta a_{\mu}$.}
\label{fig:delta_mau2-2}
\end{figure} 
For 2HDM, the two-loop Barr-Zee diagrams contribute to the $\Delta a_{\mu}$ \cite{Crivellin:2015hha,Barr:1990vd}. Where the most important contributions come from the tau, botton and top loop mediated by the scalars $H^0$ and $A^0$, since the couplings of $h$ with leptons in the considered region of parameter space are as predicted by the SM, thus the $h$ contributions are not considered in the two loop Barr-Zee contributions to $(g-2)$. This is due to the fact that in the computations of $(g-2)$, we consider the contributions arising from New Physics. For two-loop, the contribution to the anomalous magnetic moment of the muon can be obtained with the following equation \cite{Crivellin:2015hha}:
\begin{equation}
\Delta a_{\mu} =-\frac{4m_{\mu}^2}{e}\text{Re}\left(C_R\right),
\end{equation}
where $C_R$ are the Wilson coefficients, whose structure is given by~\cite{Crivellin:2013wna,Crivellin:2015hha}:
\begin{equation}
C_R=\frac{-e^3}{128\pi^4}\sum_{f_j=\tau, b, t}\frac{N_{f_j}^cQ_{f_j}^2}{m_{\mu}}\sum_{i=H^0,A^0} Y_{\overline{\mu}_{L}\mu_R}^i Y_{f_jf_j}^i\frac{m_{f_j}}{m_i^2}I_{2_i}^{(\mu)}\left(\frac{m_{f_j}^2}{m_i^2}\right),
\label{eq:m_au-2loop}
\end{equation}
where $f_j$ can be the tau, bottom and top fermions, while $N_{f_j}^c$ and $Q_{f_j}^2$ are the color and charge number of the fermion $f_j$, respectively. The loop function has the form:
\begin{eqnarray}
I_{2_{H^0}}^{(\mu)}\left(r\right)&=&\int_0^1\frac{2x(1-x)}{x(1-x)-r}\ln\left(\frac{x(1-x)}{r}\right), \\
I_{2_{A^0}}^{(\mu)}\left(r\right)&=&-\int_0^1\frac{1}{x(1-x)-r}\ln\left(\frac{x(1-x)}{r}\right).
\end{eqnarray}

The Yukawa couplings in Eq. \eqref{eq:m_au} and \eqref{eq:m_au-2loop}, according to the terms of the Lagrangian \eqref{Llep} and \eqref{Lquark} in their physical basis, are given by\footnote{The Appendix \ref{rotation} shows the rotations that were made to the Lagrangian fields \eqref{Llep} and \eqref{Lquark} and the benchmark scenario used in these matrices to go from the mass basis to the physical basis.}:
\begin{eqnarray}
&&y_{h_{\overline{\mu}_{R}\mu_L}}^{\text{SM}} = \fr{m_{\mu}}{v},  \notag \\
&&y_{h_{\overline{\mu}_{R}\mu_L}} =\fr{e^{-2 i \gamma_1}}{\sqrt{2}} \sin\alpha\left[e^{2 i \gamma_1 } \left(z_2+z_3\right) \cos ^2\theta_1+e^{i\gamma_1 } \left(z_4+z_5\right) \sin \theta_1 \cos \theta_1+\left(z_3-z_2\right) \sin ^2\theta_1\right],  \notag \\
&&y_{H_{\overline{\mu}_{R}\mu_L}^0} = -\fr{e^{-2 i \gamma_1}}{\sqrt{2}} \cos\alpha \left[e^{2 i \gamma_1} \left(z_2+z_3\right) \cos ^2\theta_1 +e^{i\gamma_1} \left(z_4+z_5\right) \sin \theta_1 \cos \theta_1-\left(z_2-z_3\right) \sin ^2\theta_1\right], \hs\, \notag\\
&&y_{A^0_{\overline{\mu}_{R}\mu_L}} = \fr{e^{-2 i \gamma_1}}{\sqrt{2}} \sin \beta \left[e^{2 i \gamma_1} \left(z_2+z_3\right) \cos ^2 \theta_1 +e^{i \gamma_1} \left(z_4+z_5\right) \sin \theta_1 \cos \theta_1 +\left(z_3-z_2\right) \sin^2 \theta_1\right].\label{eq:yuk1lopp}
\end{eqnarray}

\begin{eqnarray}
Y_{\overline{\tau}_{L}\tau_R}^{H^0}&=& \frac{\cos \alpha}{2\sqrt{2}} \Big[2 \left(z_2-z_3\right) \cos ^2\theta_1 -2 e^{-2 i \gamma_1} \left(z_2+z_3\right) \sin ^2\theta_1 -e^{-i \gamma _1} \left(z_4+z_5\right) \sin \left(2 \theta _1\right)\Big], \notag\\
Y_{\overline{\tau}_{L}\tau_R}^{A^0} &=& \fr{\sin\beta}{\sqrt{2}}\Big[e^{-2 i \gamma_1} \left(z_2+z_3\right) \sin^2\theta_1+e^{-i \gamma _1} \left(z_4+z_5\right) \sin\theta_1 \cos\theta_1+\left(z_3-z_2\right) \cos^2\theta_1\Big],\notag \\
Y_{\overline{b}_{L}b_R}^{H^0} &=& \frac{\cos\alpha}{4\sqrt{2}} \Big\{-4 e^{-i \gamma _2} \left(x_6+x_7\right) \sin\theta_2 \cos^4\theta_2-4e^{i \gamma _2} \left(x_2+x_3-x_5\right) \sin\theta_2 \cos^3\theta_2 \notag\\
&&-4e^{-2 i \gamma _2} x_1\sin^2\theta_2\left(\cos^2\theta_2+e^{i \gamma _2} \sin\theta_2\right)^2-\left(x_6+x_7\right) \sin^2\left(2 \theta_2\right) \notag\\
&&+4 \cos^3\theta_2 \left[x_2 \left(\sin^2\theta_2+\cos\theta_2\right)+x_3 \left(\sin^2\theta_2+\cos\theta_2\right)+x_4 \cos\theta_2-x_5 \sin^2\theta_2\right]\Big\}, \notag\\
Y_{\overline{b}_{L}b_R}^{A^0} &=& \fr{\sin\beta}{4\sqrt{2}} \Big\{4 e^{-i \gamma _2} \left(x_6+x_7\right) \sin\theta_2 \cos^4\theta_2+4 e^{i\gamma _2} \left(x_2+x_3-x_5\right) \sin\theta_2 \cos^3\theta_2 \notag\\
&&+4 e^{-2 i \gamma_2} x_1 \sin^2\theta_2 \left(\cos^2\theta_2+e^{i \gamma_2} \sin\theta_2\right)^2+\left(x_6+x_7\right) \sin^2\left(2 \theta _2\right)\label{eq:yuk2lopp}\\
&& -4 \cos^3\theta_2 \left[x_2 \left(\sin^2\theta_2+\cos\theta_2\right)+x_3 \left(\sin^2\theta_2+\cos\theta_2\right)+x_4 \cos\theta_2-x_5\sin^2\theta_2\right]\Big\},\notag \\
Y_{\overline{t}_{L}t_R}^{H^0} &=& \frac{\cos \alpha}{4\sqrt{2}} \Big\{4 e^{i \gamma _2} \left(w_5-w_2-w_3\right) \sin\theta_2 \cos^3\theta_2-4 e^{-2 i \gamma_2} w_1 \sin^2\theta_2 \left(\cos^2\theta_2+e^{i \gamma _2} \sin\theta_2\right)^2 \hspace{0.75cm} \notag\\
&&+4 \cos^3\theta_2 \left[w_2\left(\sin^2\theta_2 +cos\theta_2\right)+w_3 \left(\sin^2\theta_2 +cos\theta_2\right)+w_4 \cos\theta_2-w_5 \sin^2\theta_2\right]\notag\\
&&-4 e^{-i \gamma _2} \sin\theta_2 \cos^4\theta_2 \left(w_6+w_7\right)-\sin^2\left(2 \theta _2\right) \left(w_6+w_7\right)\Big\},\notag \\
Y_{\overline{t}_{L}t_R}^{A^0} &=& \frac{\sin\beta}{4\sqrt{2}} \Big\{4 e^{-i \gamma _2} \left(w_6+w_7\right) \sin\theta_2 \cos^4\theta_2+4 e^{i\gamma _2} \left(w_2+w_3-w_5\right) \sin\theta_2 \cos^3\theta_2 \notag\\
&&+4 e^{-2 i \gamma _2} w_1 \sin^2\theta_2 \left(\cos^2\theta_2 +e^{i \gamma _2} \sin\theta_2\right)^2+\left(w_6+w_7\right) \sin^2\left(2 \theta _2\right)\notag\\
&&-4 \cos^3\theta_2 \left[w_2 \left(\sin^2\theta_2+\cos\theta_2\right)+w_3 \left(\sin^2\theta_2+\cos\theta_2\right)+w_4\cos\theta_2-w_5 \sin^2\theta_2\right]\Big\},\notag
\end{eqnarray}
where the effective coupling\Revised{s} $z_j$ ($j=2,3,4,5$), $x_i$ and $w_i$ ($i=1,2,$) are given by:
\begin{eqnarray}
&&z_2=h_2v_{\eta}, z_3=h_3v_{\rho}, z_4=h_4v_{\phi_1}, z_5=h_4v_{\phi_2},\notag\\
&&x_1=h_{1d} v_{\rho}, x_2=h_{2d}v_{\rho}, x_3=h_2v_{\eta}, x_4=h_{4d}v_{\phi_1}, x_5= h_{4d}v_{\phi_2}, x_6=h_{5d}v_{\phi_1}, x_7=h_{5d}v_{\phi_2}, \notag\\
&&w_1=h_{1u}v_{\rho}, w_2=h_{2u}v_{\rho}, w_3=h_{3u}v_{\eta}, w_4=h_{4u}v_{\phi_1}, w_5= h_{4u}v_{\phi_2}, w_6=h_{5u}v_{\phi_1}, w_7=h_{5u}v_{\phi_2}. \hspace{0.75 cm}
\end{eqnarray}
Figure~\ref{fig:delta_mau2} show the values predicted by our model for $(g-2)_{\mu}$ as a function of the mass of the CP-odd neutral scalar $m_{A^0}$, considering different values for the CP-even neutral scalar ($m_{H^0}$) for the contribution to two-loop. To generate the plot of Figure \ref{fig:delta_mau2}, the mass of the CP-odd scalar was varied in the range $300\ \text{GeV}\leq m_{A^0}\leq 700\ \text{GeV}$, while the mass of the CP-even scalar was varied in the range $630\ \text{GeV}\leq m_{A^0}\leq 950\ \text{GeV}$, while the Yukawa couplings of Eqs. \eqref{eq:yuk1lopp} and \eqref{eq:yuk2lopp} were considered the following values:
\begin{align}
&y_{h_{\overline{\mu}_{R}\mu_L}} \simeq 6.46\times 10^{-2}e^{-2.87 i},\hspace{0.15 cm} y_{H_{\overline{\mu}_{R}\mu_L}^0} \simeq0.19e^{0.27 i},\hspace{0.15 cm} y_{A^0_{\overline{\mu}_{R}\mu_L}} \simeq 0.2e^{-2.87 i}, \label{eq:bestyu1}\\
&Y_{\overline{\tau}_{L}\tau_R}^{H^0}\simeq 0.79e^{0.09 i},\hspace{0.35 cm} Y_{\overline{\tau}_{L}\tau_R}^{A^0} \simeq0.83e^{-3.05 I}, \hspace{0.35 cm} Y_{\overline{b}_{L}b_R}^{H^0} \simeq 0.10e^{-3.13 i},\label{eq:bestyu2}\\
&Y_{\overline{b}_{L}b_R}^{A^0} \simeq 0.11e^{0.003 i}, \hspace{0.3 cm} Y_{\overline{t}_{L}t_R}^{H^0} \simeq 0.43e^{-3.14i}, \hspace{0.4 cm} Y_{\overline{t}_{L}t_R}^{A^0} \simeq 0.46e^{0.0035 i} .\label{eq:bestyu3}
\end{align}
These Yukawa coupling values from Eqs. \eqref{eq:bestyu1}, \eqref{eq:bestyu2} and \eqref{eq:bestyu3} are the values that best fit $\Delta a_{\mu}$, which, together with the masses of the scalars CP-even and CP-odd, whose values are:
\begin{equation}
m_{H^0}\simeq 796\ \text{GeV}, \quad m_{A^0}\simeq 387\ \text{GeV}. \label{eq:bestscalar}
\end{equation}
We obtain the following value for the deviation of the anomalous magnetic moment of the muon
\begin{equation}
\Delta a_{\mu}= 2.50\times 10^{-9}.
\end{equation}
Figure ~\ref{fig:delta_mau2-2}, we can see the values of $\Delta a_{\mu}$ as a function of the mass of the CP-odd scalar, where the same mass value of the CP-even scalar ($m_{H}=700\ \text{GeV}$) was used. We made this consideration to analyze the dominant contributions that participate in our model and allow us to obtain the values for $(g-2)_{\mu}$ shown in Fig.~\ref{fig:delta_mau2}. For this reason, different benchmarks have been taken into account, for example, the blue curve is the complete contribution of our model to two loops, which is within the upper and lower experimental limits (black horizontal lines). However, the yellow curve only considers the contribution to two loops of the tau, considering the top and bottom couplings to be zero. This benchmark can be seen in the light blue box on the right of Fig.~\ref{fig:delta_mau2-2}, which shows which couplings are considered zero in each curve. From Fig.~\ref{fig:delta_mau2-2}, we can see that the contribution of the tau is the largest. Still, it does so with negative values, limiting the values of $\Delta a_{\mu}$ and values are obtained below the lower limit, while the second highest contribution is that of the top (dark red curve), which is positive, but by itself, we obtain values for $\Delta a_{\mu}$ above the upper limit and if we add the two contributions (green curve) we obtain values for $\Delta a_{\mu}$ a little above the lower limit for values of $m_{A^0}\gtrsim 520\ \text{GeV}$. Now, if we only consider the bottom contribution (cyan curve), we can get values of $\Delta a_{\mu}$ within the experimental range, but we are ignoring the other two higher contributions, so by having all three contributions, we get values with a range of values for $\Delta a_{\mu}$ that satisfy the experimental data and with a slightly larger range of values for the mass of the CP-odd scalar.

Therefore, we find that our model can successfully accommodate the experimental values of the deviation of the anomalous magnetic moment of the muon, where all the values obtained are within the experimental range of $1\sigma$ according to Eq.~\eqref {eq:gm2}.

\section{\label{conclusion}Conclusions}

We have built a scalar singlet extension of the 2HDM with $Q_6\times Z_4\times Z_2$ symmetry that provides a successful fit of the SM fermion masses and mixing parameters. In our proposed theory, the tiny masses of the light active neutrinos are generated from a type I seesaw mechanism mediated by very heavy  right- handed Majorana neutrinos.
The feasible range of values for the Dirac phase is $\delta_{CP}\in (305.90, 348.70)^\circ$ for normal ordering and $\delta_{CP}\in (308.00, 348.00)^\circ$ for inverted ordering within the 3$\sigma$ range of experimental values. The sum of neutrino masses is obtained as $\sum m_i \in (58.03, 60.51)$ meV for normal ordering and $\sum m_i\in (98.07, 101.40)$ meV for inverted ordering which are well consistent with all the recent limits. In addition, the obtained ranges for the effective neutrino masses are $\langle m_{ee}\rangle \in (3.80, 4.38)$\, meV, $m_{\beta} \in (8.53, 9.34)\, \mbox{meV}$ for normal ordering and $\langle m_{ee}\rangle \in (47.85, 49.58) $ meV, $m_{\beta} \in (48.39, 50.09)\, \mbox{meV}$ for inverted ordering which are in agreement with the recent experimental bounds. For the quark sector, the derived results are also in agreement with the recent data on the quark masses and mixing angles. Our considered model is consistent with the muon anomalous magnetic moment.

\section*{Acknowledgments}
This research has received funding from Chilean grants ANID-Chile FONDECYT 1210378, ANID PIA/APOYO AFB230003, Milenio-ANID-ICN2019 044, PIIC program of Universidad Técnica Federico Santa María and This work was funded by Chilean grants ANID Programa de Becas Doctorado Nacional code 21212041, and from Vietnam National Foundation for Science and Technology Development (NAFOSTED) under grant number 103.01-2023.45. The authors thank Prof. Dr. Andreas Crivellin for his comments regarding the two loop level contribution of the anomalous magnetic moment of the muon, which allowed us to improve our analysis.
\appendix
	
	\section{$Q_6$ multiplication rules}\label{app}
	The $Q_6$ group has four singlets and two doublets which are, respectively, denoted by $1_{++}$, $1_{+-}$, $1_{-+}$, $1_{--}$, $2_1$ and $2_2$. Its tensor products are given by \cite{Ishi}
\begin{eqnarray}
&&\left(
	\begin{array}{c}
	\alpha_1 \\
	\alpha_2
	\end{array}
	\right)_{2_2} \otimes \left(
	\begin{array}{c}
	\beta_1 \\
	\beta_2
	\end{array}
	\right)_{2_{1}} = \left(\alpha_1 \beta_1- \alpha_2 \beta_2 \right)_{1_{+-}} \oplus
	\left(\alpha_1 \beta_1 + \alpha_2 \beta_2\right)_{1_{-+}} \oplus \left(
	\begin{array}{c}
	\alpha_1 \beta_2 \\
	\alpha_2 \beta_1
	\end{array}
	\right)_{2_{1}}, \\
&&\left(
	\begin{array}{c}
	\alpha_1 \\
	\alpha_2
	\end{array}
	\right)_{\mathbf{2}_k} \otimes \left(
	\begin{array}{c}
	\beta_1 \\
	\beta_2
	\end{array}
	\right)_{\mathbf{2}_{k}} = \left(\alpha_1 \beta_2- \alpha_2 \beta_1 \right)_{1_{++}} \oplus
	\left(\alpha_1 \beta_2 + \alpha_2 \beta_1\right)_{1_{--}} \oplus \left(
	\begin{array}{c}
	\alpha_1 \beta_1 \\
	-\alpha_2 \beta_2
	\end{array}
	\right)_{2_{k^{\prime }}}, \hspace{0.5 cm}
	\end{eqnarray}
	for $k^{\prime }\neq k$ and $k,k^{\prime }=1,2$.
	\begin{eqnarray}
	&&\hspace{-0.4 cm}\left(\alpha \right)_{1_{++}} \otimes \left(
	\begin{array}{c}
	\beta_k \\
	\beta_{-k}%
	\end{array}
	\right)_{2_k}= \left(
	\begin{array}{c}
	\alpha \beta_k \\
	\alpha \beta_{-k}%
	\end{array}
	\right)_{2_k}, \quad \left(\alpha \right)_{1_{--}} \otimes
	\left(
	\begin{array}{c}
	\beta_k \\
	\beta_{-k}%
	\end{array}
	\right)_{2_k}= \left(
	\begin{array}{c}
	\alpha \beta_k \\
	-\alpha  \beta_{-k}%
	\end{array}
	\right)_{2_k},   \\
	&&\hspace{-0.4 cm}\left(\alpha \right)_{1_{+-}} \otimes \left(
	\begin{array}{c}
	\beta_k \\
	\beta_{-k}%
	\end{array}
	\right)_{2_k}= \left(
	\begin{array}{c}
	\alpha \beta_{-k} \\
	\alpha \beta_{k}%
	\end{array}
	\right)_{2_{3-k}}, \quad \left(\alpha \right)_{1_{-+}} \otimes
	\left(
	\begin{array}{c}
	\beta_{k} \\
	\beta_{-k}%
	\end{array}
	\right)_{2_k}= \left(
	\begin{array}{c}
	\alpha \beta_{-k} \\
	- \alpha \beta_{k}%
	\end{array}
	\right)_{2_{3-k}},\\
&&\hspace{-0.4 cm}1_{s_1s_2} \otimes 1_{s^{\prime }_1 s^{\prime }_2} =
	1_{s^{\prime \prime }_1s^{\prime \prime }_2},
	\end{eqnarray}
	where $s^{\prime \prime }_1=s_1s^{\prime }_1$ and $s^{\prime \prime
	}_2=s_2s^{\prime }_2$.
\section{\label{Preventedterms} Prevented terms}
\vspace{-0.25 cm}
 \begin{table}[h]
  \vspace{-0.25 cm}
  \caption{\label{PreventedtermsT} Prevented terms.}
  \vspace{0.5 cm}
 \begin{tabular}{|c|c|c|c|c|c|c|c|} \hline
\multirow{2}{5.5cm}{\hfill Prevented terms 
\hfill } & Prevented \\
\cline{3-1}   & by \\  \hline

$(\overline{\psi}_{1L} l_{1R})_{1_{++}} (S\rho^*)_{1_{++}}, (\overline{\psi}_{L} \psi^C_{L})_{1_{--}} \widetilde{H}^2,
(\overline{\psi}_{1L} \nu_{1R})_{1_{++}} \widetilde{S}, (\bar{\psi}_{L} \nu_{1R})_{2_{2}} (\widetilde{H} \phi)_{2_{1}},
$&\multirow{11}{0.45 cm}{$Q_6$}  \\
$ (\bar{\nu}^c_{1R} \nu_{1R})_{1_{++}} (\rho\eta)_{\underline{1}_{--}},
(\bar{\nu}^c_{1R} \nu_{1R})_{1_{++}} (\rho^*\eta^*)_{\underline{1}_{--}},
(\bar{\nu}^c_{1R} \nu_{1R})_{1_{++}} (\rho\phi)_{\underline{2}_{1}}, $&\\
$ (\bar{\nu}^c_{1R} \nu_{1R})_{1_{++}} (\rho^*\phi^*)_{\underline{2}_{1}},
(\bar{\nu}^c_{1R} \nu_{1R})_{1_{++}} (\eta\phi)_{\underline{2}_{1}},
(\bar{\nu}^c_{1R} \nu_{1R})_{1_{++}} (\eta^*\phi^*)_{\underline{2}_{1}}, $&\\
$(\bar{\nu}^c_{1R} \nu_{1R})_{1_{++}} (\phi^2)_{\underline{1}_{--}},
(\bar{\nu}^c_{R} \nu_{R})_{1_{++}} (\rho\eta)_{\underline{1}_{--}},
(\bar{\nu}^c_{R} \nu_{R})_{1_{++}} (\rho^*\eta^*)_{\underline{1}_{--}},
(\bar{\nu}^c_{R} \nu_{R})_{1_{++}} (\rho\phi)_{\underline{2}_{1}},  $&\\
$ (\bar{\nu}^c_{R} \nu_{R})_{1_{++}} (\rho^*\phi^*)_{\underline{2}_{1}},
(\bar{\nu}^c_{R} \nu_{R})_{1_{++}} (\eta\phi)_{\underline{2}_{1}},
(\bar{\nu}^c_{R} \nu_{R})_{1_{++}} (\eta^*\phi^*)_{\underline{2}_{1}},
(\bar{\nu}^c_{R} \nu_{R})_{1_{++}} (\phi^2)_{\underline{1}_{--}}; $&\\
$(\bar{Q}_{1 L}  u_{1 R})_{1_{+-}}(\widetilde{H}\eta^*)_{1_{-+}},
(\bar{Q}_{L}  u_{R})_{1_{+-}}(\widetilde{H}\eta^*)_{1_{-+}},
(\bar{Q}_{L}  u_{R})_{1_{-+}}(\widetilde{H}\rho^*)_{1_{+-}}, $& \\
$(\bar{Q}_{L}  u_{R})_{1_{-+}}(\widetilde{H}\phi^*)_{2_{1}},
(\bar{Q}_{L}  u_{R})_{2_{1}}(\widetilde{H}\rho^*)_{1_{+-}},
(\bar{Q}_{L}  u_{R})_{2_{1}}(\widetilde{H}\eta^*)_{1_{-+}},
(\bar{Q}_{1L}  u_{R})_{2_{1}}(\widetilde{H}\rho^*)_{1_{+-}}, $&\\
$(\bar{Q}_{1L}  u_{R})_{2_{1}}(\widetilde{H}\eta^*)_{1_{-+}},
(\bar{Q}_{L}  u_{2R})_{2_{1}}(\widetilde{H}\rho^*)_{1_{+-}},
(\bar{Q}_{L}  u_{2R})_{2_{1}}(\widetilde{H}\eta^*)_{1_{-+}}, $ &\\
$(\bar{Q}_{1L} d_{1R})_{1_{+-}} (H\eta^*)_{1_{+-}},
(\bar{Q}_{1L} d_{1R})_{1_{+-}} (H\phi^*)_{2_{1}},
(\bar{Q}_{L} d_{R})_{1_{-+}} (H\rho^*)_{1_{+-}},$&\\
$(\bar{Q}_{L} d_{R})_{1_{-+}} (H\phi^*)_{2_{1}},
(\bar{Q}_{1L} d_{R})_{2_{1}} (H\rho^*)_{1_{+-}},
(\bar{Q}_{L} d_{1R})_{2_{1}} (H\rho^*)_{1_{+-}}, $&\\
$(\bar{Q}_{L} d_{R})_{2_{1}} (H\rho^*)_{1_{+-}},
(\bar{Q}_{1L} d_{R})_{2_{1}} (H\eta^*)_{1_{-+}},
(\bar{Q}_{1L} d_{R})_{2_{1}} (H\eta^*)_{1_{-+}},
(\bar{Q}_{L} d_{R})_{2_{1}} (H\eta^*)_{1_{-+}} $&\\\hline
$(\overline{\psi}_{1L} l_{1R})_{1_{++}} (S\rho)_{1_{++}},
(\overline{\psi}_{L} l_{R})_{1_{+-}} (H\rho^*)_{1_{+-}},
(\overline{\psi}_{L} l_{R})_{1_{-+}} (H\eta^*)_{1_{-+}},
(\overline{\psi}_{L} l_{R})_{2_{1}} (H\phi^*)_{2_{1}},$& \multirow{9}{0.4 cm}{$Z_4$} \\
$(\overline{\psi}_{1L} \psi^C_{1L})_{1_{++}} \widetilde{S}^2,
(\overline{\psi}_{L} \psi^C_{L})_{1_{++}}\widetilde{S}^2, (\bar{\psi}_{1L} \nu_{R})_{2_{1}} (\widetilde{H} \phi^*)_{2_{1}},
(\bar{\psi}_{L} \nu_{1R})_{2_{2}} (\widetilde{S} \phi)_{2_{1}},
$& \\
$(\bar{\psi}_{L} \nu_{1R})_{2_{2}} (\widetilde{S} \phi^*)_{2_{1}},
(\bar{\nu}^c_{1R} \nu_{1R})_{1_{++}} (\phi\phi^*)_{\underline{1}_{++}},
(\bar{\nu}^c_{R} \nu_{R})_{1_{++}} (\phi\phi^*)_{\underline{1}_{++}},
(\bar{\nu}^c_{R} \nu_{R})_{1_{--}} (S\widetilde{S} )_{\underline{1}_{--}}, $&\\
$(\bar{\nu}^c_{R} \nu_{R})_{1_{--}} (\rho\rho^*)_{\underline{1}_{--}},
(\bar{\nu}^c_{R} \nu_{R})_{1_{--}} (\eta\eta^*)_{\underline{1}_{--}},
(\bar{\nu}^c_{R} \nu_{R})_{1_{--}} (\phi\phi^*)_{\underline{1}_{--}},
(\bar{\nu}^c_{R} \nu_{R})_{2_{2}} (\rho^*\phi)_{\underline{2}_{2}},$&\\
$(\bar{\nu}^c_{R} \nu_{R})_{2_{2}} (\rho\phi^*)_{\underline{2}_{2}},
(\bar{\nu}^c_{R} \nu_{R})_{2_{2}} (\eta^*\phi)_{\underline{2}_{2}},
(\bar{\nu}^c_{R} \nu_{R})_{2_{2}} (\eta\phi^*)_{\underline{2}_{2}};
(\bar{Q}_{1 L}  u_{1 R})_{1_{+-}}(\widetilde{H}\eta)_{1_{+-}}, $&\\
$(\bar{Q}_{L}  u_{R})_{1_{+-}}(\widetilde{H}\eta)_{1_{+-}},
(\bar{Q}_{L}  u_{R})_{1_{-+}}(\widetilde{H}\rho)_{1_{-+}},
(\bar{Q}_{L}  u_{R})_{2_{1}}(\widetilde{H}\phi)_{2_{1}},
(\bar{Q}_{1L}  u_{R})_{1_{+-}}(\widetilde{H}\eta)_{1_{+-}}, $& \\
$(\bar{Q}_{1L}  u_{R})_{1_{-+}}(\widetilde{H}\rho)_{1_{-+}},
(\bar{Q}_{1L}  u_{R})_{2_{1}}(\widetilde{H}\phi)_{2_{1}},
(\bar{Q}_{L}  u_{1R})_{1_{+-}}(\widetilde{H}\eta)_{1_{+-}},
(\bar{Q}_{L}  u_{1R})_{1_{-+}}(\widetilde{H}\rho)_{1_{-+}}, $&\\
$(\bar{Q}_{L}  u_{1R})_{2_{1}}(\widetilde{H}\phi)_{2_{1}},
(\bar{Q}_{1 L}  d_{1 R})_{1_{+-}}(H\eta)_{1_{+-}},
(\bar{Q}_{1L} d_{1R})_{1_{-+}} (H\rho)_{1_{-+}},$ &\\

$(\bar{Q}_{1L} d_{R})_{2_{1}} (H\phi)_{2_{1}},
(\bar{Q}_{L} d_{1R})_{2_{1}} (H\phi)_{2_{1}},
(\bar{Q}_{L} d_{R})_{2_{1}} (H\phi)_{2_{1}}$&\\

 \hline
$(\overline{\psi}_{1L} l_{R})_{2_1} (H\phi)_{2_1}, (\overline{\psi}_{L} l_{1R})_{2_2} (S\phi^*)_{2_2},
(\bar{\psi}_{L} \nu_{R})_{1_{-+}} (\widetilde{H} \rho)_{1_{-+}}, (\bar{\psi}_{L} \nu_{R})_{1_{+-}} (\widetilde{H} \eta)_{1_{+-}},$&\multirow{2}{0.4 cm}{$Z_2$}  \\
$(\bar{\psi}_{L} \nu_{R})_{2_{1}} (\widetilde{H} \phi)_{2_{1}},
(\bar{Q}_{1 L}  d_{1 R})_{1_{+-}}S$&\\ \hline
\end{tabular}
\vspace{-1.5 cm}
\end{table}
\newpage
\section{\label{Higgs} Scalar potential and stability condition}
The potential invariant under $\Gamma$ gets the following form\footnote{It is noted that $(\phi^*\phi)_{\underline{1}_{--}}=0,\, (\phi^*\phi)_{\underline{1}_{--}}(\phi^*\phi)_{\underline{1}_{--}}=0,\, \big[(\phi^* \phi)_{\underline{2}_2}(\phi^* \phi)_{\underline{2}_2}\big]_{1_{++}}=0$, $V(H,\rho)=V(H,\eta)=V(S,\rho)=V(S,\eta)=V(S,\phi)=V(\rho,\phi)=V(\eta,\phi)=0$ due to the VEV alignment of $\phi$ in Eq. (\ref{scalarvev}) and the tensor product of $Q_6$ group. Furthermore, other Yukawa terms with three and four different scalars are all not invariant under one or some of symmetries of $\Gamma$; thus, they were not included in the expressions of $V_{\mathrm{scalar}}$.}:
\bea V_{\mathrm{scalar}}&=& V(H)+V(S)+
V(\rho)+V(\eta)+V(\phi)+ V(H, S)+V(H, \phi)+V(\rho,\eta) + V_{\mathrm{tri}},
\label{Higgspoten}\eea
where
\bea
&&\hspace{-0.25 cm}V(H)=\mu^2_H H^\+H +\lambda^H (H^\+H)_{\underline{1}_{++}}(H^\+H)_{\underline{1}_{++}}, \hs V(S)=V(H\rightarrow S),\crn
&&\hspace{-0.25 cm}V(\rho)=\lambda^\rho (\rho^* \rho)_{\underline{1}_{--}}(\rho^* \rho)_{\underline{1}_{--}},\,\, V(\eta)=V(\rho\rightarrow \eta), 
\,\, V(\phi)=\mu^2_\phi (\phi^*\phi)_{\underline{1}_{++}} +\lambda^\phi(\phi^* \phi)^2_{\underline{1}_{++}}, 
\crn
&&\hspace{-0.25 cm}V(H,S)=\lambda^{HS} \big[(S^\+ H)_{\underline{1}_{++}}^2+h.c.\big],
V(H, \phi)=\lambda^{H\phi}_1 (H^\+H)_{1_{++}}(\phi^*\phi)_{1_{++}}+\lambda^{H\phi}_2(H^\+\phi)_{2_{1}}(\phi^*H)_{2_{1}}, \crn
&&\hspace{-0.25 cm}V(\rho,\eta)=\lambda^{\rho\eta}_1(\rho^*\rho)_{\underline{1}_{--}}(\eta^* \eta)_{\underline{1}_{--}}+\lambda^{\rho\eta}_2(\rho^* \eta)_{\underline{1}_{++}}(\eta^* \rho)_{\underline{1}_{++}}, \,V_{\mathrm{tri}}=\lambda^{H\rho\eta} (H^\+ H)_{\underline{1}_{++}}\big[(\rho^*\eta)_{\underline{1}_{++}}+h.c\big] \hspace{0.5 cm} \crn
&&\hspace{1.1 cm}+ \lambda^{\rho\eta\phi}_1(\rho^*\eta+\rho\eta^*)_{\underline{1}_{++}}(\phi^*\phi)_{\underline{1}_{++}}
+\lambda^{\rho\eta\phi}_2(\rho\eta)_{\underline{1}_{--}}(\phi^2)_{\underline{1}_{--}}
+\lambda^{\rho\eta\phi}_3(\rho^*\eta^*)_{\underline{1}_{--}}(\phi^{*2})_{\underline{1}_{--}}, \label{HFs}
\eea
We will show that the VEV alignment in Eq. (\ref{scalarvev}) satisfies the
minimization condition of the scalar potential by assuming that all the VEVs are real. In this case, $\fr{\partial V_{\mathrm{scalar}}}{\partial v^*_\alpha} =\fr{\partial V_{\mathrm{scalar}}}{\partial v_\alpha},\, \fr{\partial^2 V_{\mathrm{scalar}}}{\partial v^{*2}_\alpha}=\fr{\partial^2 V_{\mathrm{scalar}}}{\partial v^2_\alpha}\, (\alpha =H, S, \rho,\eta,\phi)$, and the
minimization condition reduce to:
\bea
&&\mu^2_H + 2 \lambda^{H} v_H^2 + 2 \big(\lambda^{H\phi}_2-\lambda^{H\phi}_1\big) v_\phi^2 + \lambda^{H\rho\eta} v_\eta v_\rho +
 4 \lambda^{HS} v_S^2=0, \\
&&\mu^2_S + 4 \lambda^{HS} v_H^2 + 2 \lambda^{S} v_S^2=0, \\
&&\lambda^{H\rho\eta} v_\eta v_H^2 -
 2 \big(2 \lambda^{\rho\eta\phi}_1 - \lambda^{\rho\eta\phi}_2 + \lambda^{\rho\eta\phi}_3\big) v_\eta v_\phi^2 +
 2 \big(\lambda^{\rho\eta}_1 + \lambda^{\rho\eta}_2\big) v_\eta^2 v_\rho + 4 \lambda^{\rho} v_\rho^3=0,\crn
 &&v_{\rho} \left[2 v_{\eta} v_{\rho} \big(\lambda^{\rho\eta}_1+\lambda^{\rho\eta}_2\big)-2 v_{\phi}^2 \big(2 \lambda^{\rho\eta\phi}_1-\lambda^{\rho\eta\phi}_2+\lambda^{\rho\eta\phi}_3)+\lambda^{H\rho\eta} v_H^2\right]+4 \lambda^{\eta} v_{\eta}^3=0, \crn
&&v_H^2 \big(\lambda^{H\phi}_2-\lambda^{H\phi}_1\big)+v_{\eta} v_{\rho} \big(-2 \lambda^{\rho\eta\phi}_1+\lambda^{\rho\eta\phi}_2-\lambda^{\rho\eta\phi}_3\big)+4 \lambda^{\phi} v_{\phi}^2-\mu^2_\phi=0,\crn
&&\mu^2_\phi-v_H^2 \big(\lambda^{H\phi}_2-\lambda^{H\phi}_1\big)-v_{\eta} v_{\rho} \big(\lambda^{\rho\eta\phi}_2-\lambda^{\rho\eta\phi}_3-2 \lambda^{\rho\eta\phi}_1\big)-4 \lambda^{\phi} v_{\phi}^2=0,
\eea
and
\bea
&&\lambda^{H} v^2_H>0, \hs \lambda^{S} v^2_S>0,\hs \big(\lambda^{\rho\eta}_1 + \lambda^{\rho\eta}_2) v_\eta^2 + 6 \lambda^{\rho} v_\rho^2>0,  \hs \lambda^{\phi} v_\phi^2 >0. \label{ineq}
\eea
Using the VEVs of the scalar fields in Eqs. (\ref{vSvH}) and (\ref{vevscales}), with the help of expression (\ref{ineq}), we obtain the following conditions for the scalar potential stability condition:
\bea
  &&\lambda^{H}>0, \hs \lambda^{S}>0, \hs \lambda^{\phi}>0, \hs
\lambda^{\rho}>-\frac{\big(\lambda^{\rho\eta}_1+\lambda^{\rho\eta}_2\big) v_\eta^2}{6 v_\rho^2}. \label{condit3}
  \eea

\section{\label{rotation} Base Rotation}
Here, we will see the rotations from the mass base to the physical base, which have been used to perform the analysis of $(g-2)_{\mu}$ in section \ref{anomal}.

For the scalar fields $H^0$ and $A^0$ we have the following rotation:
\begin{eqnarray}
\begin{pmatrix}
\sigma_H \\
\sigma_S
\end{pmatrix}&=&\begin{pmatrix}
\sin\alpha & -\cos\alpha \\
-\cos\alpha & -\sin\alpha
\end{pmatrix}
\begin{pmatrix}
h\\
H^0
\end{pmatrix},\\[10pt]
\begin{pmatrix}
\eta_H \\
\eta_S
\end{pmatrix}&=&\begin{pmatrix}
\cos\beta & \sin\beta \\
\sin\beta & -\cos\beta
\end{pmatrix}
\begin{pmatrix}
G^0\\
A^0
\end{pmatrix}.
\end{eqnarray}
Therefore, we have to
\begin{align}
\sigma_H&= h\sin\alpha -H^0\cos\alpha, \hs
\sigma_S= -h\cos\beta -H^0\sin\beta.
\end{align}
From Eq.~\eqref{Mclep} for the charged lepton sector, we have only one rotation in the $1-2$ plane, so we have:
\begin{eqnarray}
\begin{pmatrix}
l_{L,R}^{(1)} \\
l_{L,R}^{(2)} \\
l_{L,R}^{(3)}
\end{pmatrix}&=& \begin{pmatrix}
1 & 0 & 0 \\
 0 & \cos\theta _1 & -e^{-i \gamma _1} \sin\theta _1 \\
 0 & e^{i \gamma _1} \sin\theta_1 & \cos\theta_1
\end{pmatrix}
\begin{pmatrix}
\psi_{L,R}^{(1)} \\
\psi_{L,R}^{(2)} \\
\psi_{L,R}^{(3)}
\end{pmatrix}.
\end{eqnarray}
where we get
\begin{align}
l_{L,R}^{(1)}&= \psi_{L,R}^{(1)}, \\
l_{L,R}^{(2)}&= \psi_{L,R}^{(2)}\cos\theta_1 -\psi_{L,R}^{(3)}e^{-i \gamma _1} \sin\theta_1,\\
l_{L,R}^{(3)}&= \psi_{L,R}^{(2)}e^{i \gamma _1} \sin\theta_1 +\psi_{L,R}^{(3)}\cos\theta_1.
\end{align}
In the quark sector, we have to perform a rotation in the three planes so that we can use the following matrix:
\begin{equation}
R_q=R_{1q}\cdot  R_{2q}\cdot R_{3q},
\end{equation}
where,
\begin{align}
R_{1q}&= \begin{pmatrix}
\cos \theta _k & -e^{-i \gamma _k} \sin \theta _k & 0 \\
 e^{i \gamma _k} \sin \theta _k & \cos \theta _k & 0 \\
 0 & 0 & 1
\end{pmatrix},\quad R_{2q}= \begin{pmatrix}
\cos \theta_k & 0 & -e^{-i \gamma _k} \sin \theta_k \\
 0 & 1 & 0 \\
 e^{i \gamma _k} \sin \theta_k & 0 & \cos \theta_k
\end{pmatrix},\notag\\
R_{3q}&= \begin{pmatrix}
1 & 0 & 0 \\
 0 & \cos \theta_k & e^{i \gamma _k} \sin \theta_k \\
 0 & -e^{-i \gamma _k} \sin \theta_k & \cos \theta_k
\end{pmatrix},
\end{align}
which give us the following matrix:
\begin{equation}
R_q=\begin{pmatrix}
\cos ^2\theta_k & e^{-2 i \gamma _k} \sin ^2\theta_k  \cos \theta_k-e^{-i \gamma _k} \sin \theta_k \cos \theta_k & -\sin ^2\theta_k-e^{-i \gamma _k} \sin  \theta_k \cos ^2\theta_k \\
 e^{i \gamma _k} \sin \theta_k \cos \theta_k & \cos^2\theta_k+e^{-i \gamma _k} \sin ^3\theta_k & -\sin^2\theta_k \cos \theta_k+e^{i \gamma _k} \sin\theta_k \cos \theta_k \\
 e^{i \gamma _k} \sin \theta_k & -e^{-i \gamma _k} \sin\theta_k \cos \theta_k & \cos ^2\theta_k
\end{pmatrix},
\label{eq:Rq}
\end{equation}
where $q=u,d$ while $k=2$ for the up quarks and $k=3$ for the down quarks. With the matrix \eqref{eq:Rq} we will obtain:
\begin{eqnarray}
\begin{pmatrix}
q_{L,R}^{(1)} \\
q_{L,R}^{(2)}\\
q_{L,R}^{(3)}
\end{pmatrix}= R_q\begin{pmatrix}
Q_{L,R}^{(1)} \\
Q_{L,R}^{(2)} \\
Q_{L,R}^{(3)}
\end{pmatrix}.
\end{eqnarray}
Therefore, we have:
\begin{align}
q_{L,R}^{(1)}&= Q_{L,R}^{(1)}\cos ^2\theta_k +Q_{L,R}^{(2)}\left( e^{-2 i \gamma _k} \sin ^2\theta_k  \cos \theta_k-e^{-i \gamma _k} \sin \theta_k \cos \theta_k\right) \notag\\
 &-Q_{L,R}^{(3)} \left(\sin ^2\theta_k +e^{-i \gamma _k} \sin  \theta_k \cos ^2\theta_k\right), \notag\\
q_{L,R}^{(2)}&= Q_{L,R}^{(1)} e^{i \gamma _k} \sin \theta_k \cos \theta_k +Q_{L,R}^{(2)} \left(\cos^2\theta_k+e^{-i \gamma _k} \sin ^3\theta_k\right), \notag\\
 &+Q_{L,R}^{(3)}\left(-\sin^2\theta_k \cos \theta_k+e^{i \gamma _k} \sin\theta_k \cos \theta_k\right), \notag\\
q_{L,R}^{(3)}&= Q_{L,R}^{(1)}e^{i \gamma _k} \sin \theta_k - Q_{L,R}^{(2)}e^{-i \gamma _k} \sin\theta_k \cos \theta_k +Q_{L,R}^{(3)} \cos ^2\theta_k.
\end{align}


\newpage
\begin{thebibliography}{99}
\bibitem{Feruglio15} F. Feruglio, \emph{Pieces of the flavour puzzle}, Eur. Phys. J. C 75 (2015) 373. https://doi.org/10.1140/epjc/s10052-015-3576-5.
\bibitem{Novichkov21} P. P. Novichkov, J. T. Penedo, S. T. Petcov, \emph{Fermion Mass Hierarchies, Large Lepton Mixing and Residual Modular Symmetries}, J. High Energ. Phys. 2021, 206 (2021). https://doi.org/10.1007/JHEP04(2021)206.
\bibitem{PDG2022} R. L. Workman \emph{et al.} (Particle Data Group), \emph{Review of Particle Physics}, Prog. Theor. Exp. Phys. 2022 (2022) 083C01. https://doi.org/10.1093/ptep/ptac097.

\bibitem{Salas2021} P. F. de Salas \emph{et al.}, \emph{2020 Global reassessment of the neutrino oscillation picture},  J. High Energ. Phys. 2021, 71 (2021). https://doi.org/10.1007/JHEP02(2021)071. 


\bibitem{King:2013eh}
S.~F. King and C.~Luhn, \emph{Neutrino Mass and Mixing with Discrete Symmetry},
Rept. Prog. Phys. 76 (2013) 056201. http://dx.doi.org/10.1088/0034-4885/76/5/056201.

\bibitem{Altarelli:2010gt}
G.~Altarelli and F.~Feruglio, \emph{Discrete Flavor Symmetries and Models of
  Neutrino Mixing},
 Rev. Mod. Phys.  82(2010) 2701. http://dx.doi.org/10.1103/RevModPhys.82.2701.
\bibitem{VLA2020S3} V.V. Vien, H.N. Long, and A.E. Cárcamo Hernández, \emph{$U(1)_{B-L}$ extension of the standard model with $S_3$
 symmetry}, Eur. Phys. J. C 80 (2020) 725. https://doi.org/10.1140/epjc/s10052-020-8318-7.


\bibitem{King:2015aea}
S.~F. King, \emph{Models of Neutrino Mass, Mixing and CP Violation},
J. Phys. G 42 (2015) 123001. http://dx.doi.org/10.1088/0954-3899/42/12/123001.
\bibitem{Q61} K. S. Babu and J. Kubo, \emph{Dihedral families of quarks, leptons and Higgses},
Phys.Rev. D 71 (2005) 056006.
https://doi.org/10.1103/PhysRevD.71.056006. 

\bibitem{Q62} Y. Kajiyama, E. Itou and J. Kubo, \emph{Nonabelian discrete family symmetry to soften the SUSY flavor problem and to suppress proton decay},
Nucl. Phys. B 743 (2006) 74.
https://doi.org/10.1016/j.nuclphysb.2006.02.042. 
\bibitem{Q63} Y. Kajiyama, 
J. High Eenergy Phys. 04 (2007) 007, arXiv:hep-ph/0702056.
\bibitem{Q64} K. Babu, K. Kawashima and J. Kubo, \emph{Variations on the Supersymmetric $Q_6$ Model of Flavor},
Phys.Rev. D 83 (2011) 095008.
https://doi.org/10.1103/PhysRevD.83.095008.
\bibitem{Q65} Takeshi Araki and Y. F. Li, \emph{$Q_6$ flavor symmetry model for the extension of the minimal standard model by three right-handed sterile neutrinos}, Phys. Rev. D 85 (2012) 065016. https://doi.org/10.1103/PhysRevD.85.065016. 

\bibitem{Q66} J. Kubo, \emph{Super Flavorsymmetry with Multiple Higgs Doublets}, Fortsch.Phys. 61 (2013) 597.
https://doi.org/10.1002/prop.201200119. 

\bibitem{Q67} Juan Carlos Gómez-Izquierdo, F. Gonzalez-Canales, M. Mondragón, \emph{On $Q_6$ flavor symmetry and the breaking of $\mu\leftrightarrow\tau$ symmetry}, Int. J. Mod. Phys. A 32 (2017) 1750171.
https://doi.org/10.1142/S0217751X17501718.

\bibitem{Q68} V. V. Vien, \emph{B-L extension of the standard model with $Q_6$ symmetry}, Nucl. Phys. B 956 (2020) 115015. https://doi.org/10.1016/j.nuclphysb.2020.115015.
\bibitem{Pokorski05} S. Pokorski, \emph{Phenomenological guide to physics beyond the Standard Model}, arXiv:hep-ph/0502132.
\bibitem{Chunjhep12} E. J. Chun, H. M. Lee, and P. Sharma, \emph{Vacuum stability, perturbativity, EWPD and Higgs-to-diphoton rate in type II seesaw models}, J. High Energ. Phys. 2012 (2012) 106. https://doi.org/10.1007/JHEP11(2012)106. arXiv:1209.1303 [hep-ph].
\bibitem{cutoffscal21} K. Aoki, T. Q. Loc, T. Noumi, and J. Tokuda, \emph{Is the Standard Model in the Swampland? Consistency Requirements from Gravitational Scattering}, Phys. Rev. Lett. 127 (2021) 091602. https://doi.org/10.1103/PhysRevLett.127.091602. arXiv: 2104.09682.
\bibitem{Asakajhep11} Takehiko Asaka, Shintaro Eijima, Hiroyuki Ishida, \emph{Mixing of Active and Sterile Neutrinos},  J. High Energ. Phys. 2011 (2011) 11. https://doi.org/10.1007/JHEP04(2011)011. 	arXiv:1101.1382 [hep-ph].
\bibitem{Blennowjhep10} M. Blennow, E. Fernandez-Martinez, J. Lopez-Pavon and J. Menendez,
  \emph{Neutrinoless double beta decay in seesaw models}, J. High Energ. Phys. 2010 (2010) 96. https://doi.org/10.1007/JHEP07(2010)096. arXiv:1005.3240 [hep-ph].
\bibitem{Benes2005} P. Benes, A. Faessler, F. Simkovic and S. Kovalenko, \emph{Sterile neutrinos in neutrinoless double beta decay}, Phys. Rev. D {\bf 71} (2005) 077901. https://doi.org/10.1103/PhysRevD.71.077901. arXiv:hep-ph/0501295.

\bibitem{Bonilla:2021ize} C.~Bonilla, A.~E.~C\'arcamo Hern\'andez, J.~Gon\c{c}alves, F.~F.~Freitas, A.~P.~Morais and R.~Pasechnik,
\emph{Collider signatures of vector-like fermions from a flavor symmetric model}, JHEP \textbf{01}, 154 (2022), doi:10.1007/JHEP01(2022)154.


\bibitem{Wang2hdm} Lei Wang, Jin Min Yang, Mengchao Zhang, Yang Zhang, \emph{Revisiting lepton-specific 2HDM in light of muon $g-2$ anomaly}, Phys.Lett. B788 (2019) 519.
https://doi.org/10.1016/j.physletb.2018.11.045. arXiv:1809.05857 [hep-ph].
\bibitem{Cao2hdm} Junjie Cao, Peihua Wan, Lei Wu, Jin Min Yang, \emph{Lepton-Specific Two-Higgs Doublet Model: Experimental Constraints and Implication on Higgs Phenomenology}, Phys.Rev.D 80 (2009)071701.
https://doi.org/10.1103/PhysRevD.80.071701. 	arXiv:0909.5148 [hep-ph].
\bibitem{Chun2hdm} Eung Jin Chun, Tanmoy Mondal, \emph{Explaining g-2 anomalies in two Higgs doublet model with vector-like leptons}, J. High Energ. Phys. 2020(2020)77 . https://doi.org/10.1007/JHEP11(2020)077. arXiv:2009.08314 [hep-ph].

\bibitem{Chun2hdm15} Eung Jin Chun, \emph{The Muon g–2 in two-Higgs-doublet models}, EPJ Web of Conferences 118, 01006 (2016).
https://doi.org/10.1007/s12043-016-1254-2. 	arXiv:1511.05225 [hep-ph].
\bibitem{Abe2hdm} T. Abe, R. Sato and K. Yagyu, \emph{Muon specific two-Higgs-doublet model}, J. High Energ. Phys. 2017 (2017) 12. https://doi.org/10.1007/JHEP07(2017)012. 	arXiv:1705.01469 [hep-ph]







\bibitem{Ishi} H. Ishimori \emph{et al.,} 
\emph{An introduction to non-Abelian discrete symmetries for particle physicists}, Lect. Notes Phys. 858 (2012) 1.



\bibitem{RoyChoudhury} S.  R. Choudhury, Shouvik and Hannestad, Steen, \emph{Updated results on neutrino mass and mass hierarchy from cosmology with Planck 2018 likelihoods}, JCAP 2007 (2020) 037. https://doi.org/10.1088/1475-7516/2020/07/037.
\bibitem{Tanseri22} I. Tanseri \emph{et al.}, \emph{Updated neutrino mass constraints from galaxy clustering and CMB lensing-galaxy cross-correlation measurements}, 	JHEAp 36 (2022) 1. https://doi.org/10.1016/j.jheap.2022.07.002.
\bibitem{Stockerprd21} P. Stöcker et al. [GAMBIT Cosmology Workgroup], \emph{Strengthening the bound on the mass of the lightest neutrino with terrestrial and cosmological experiments}, Phys. Rev. D 103 (2021) 123508. https://doi.org/10.1103/PhysRevD.103.123508. 
\bibitem{Alamprd21} Shadab Alam \emph{et al.}, \emph{Completed SDSS-IV extended Baryon Oscillation Spectroscopic Survey: Cosmological implications from two decades of spectroscopic surveys at the Apache Point Observatory}, Phys. Rev. D 103 (2021) 083533.
https://doi.org/10.48550/arXiv.2007.08991.



\bibitem{GuillermoJHEP21} G. F. Abellán, 
Z. Chacko, A. Dev, P. Du, V. Poulin and Y. Tsai, \emph{Improved cosmological constraints on the neutrino mass and lifetime},
    \emph{J. High Energ. Phys.} 2022 (2022) 76. https://doi.org/10.1007/JHEP08(2022)076.
\bibitem{Adame2024arxiv} A. G. Adame et al. (DESI Collaboration), \emph{DESI 2024 VI: Cosmological Constraints from the Measurements of Baryon Acoustic Oscillations}, arXiv:2404.03002 [astro-ph.CO].
\bibitem{Jimenezjcap22} R. Jimenez, C. Pena-Garay, K. Short, F. Simpson and L. Verde, \emph{Neutrino masses and mass hierarchy: Evidence for the normal hierarchy}, JCAP 09 (2022) 006. https://doi.org/10.1088/1475-7516/2022/09/006. 

\bibitem{KamLAND16} A. Gando \emph{et al.} (KamLAND-Zen Collaboration), \emph{Search for Majorana Neutrinos Near the Inverted Mass Hierarchy Region with KamLAND-Zen},  Phys.Rev.Lett. 117 (2016) 082503. https://doi.org/10.1103/PhysRevLett.117.082503. 
\bibitem{GERDA19} M. Agostini \emph{et al.} (GERDA Collaboration), \emph{Probing Majorana neutrinos with double-$\beta$ decay}, Science 365 (2019) 1445. DOI: 10.1126/science.aav8613.
\bibitem{CUORE20} D. Adams \emph{et al.} (CUORE collaboration), \emph{Improved Limit on Neutrinoless Double-Beta Decay in $^{130}Te$ with CUORE}, Phys.Rev.Lett. 124 (2020) 122501. https://doi.org/10.1103/PhysRevLett.124.122501.


\bibitem{mbet3constraint} Jun Cao \emph{et al}, \emph{Towards the meV limit of the effective neutrino mass in neutrinoless double-beta decays}, Chinese Phys. C 44 (2020) 031001. https://doi.org/10.1088/1674-1137/44/3/031001.

\bibitem{Aker22n} M. Aker \emph{et al.} (The KATRIN Collaboration), \emph{Direct neutrino-mass measurement with sub-electronvolt sensitivity}, Nat. Phys. 18 (2022) 160. https://doi.org/10.1038/s41567-021-01463-1.


\bibitem{JCao2020mev} J. Cao \emph{et al.}, \emph{Towards the meV limit of the effective neutrino mass in neutrinoless double-beta decays}, Chinese Phys. C 44 (2020) 031001.

\bibitem{Renk2002proceeding} B. Renk, \emph{Status of the Cabibbo-Kobayashi-Maskawa quark-mixing matrix}, Proceedings, International Workshop, Heidelberg, Germany, 19-20 Sep. 2002, 3-8.
\bibitem{Cruz24arxiv} E. P. Cruz, A. O. Bouzas, F. Larios, \emph{Probing $|V_{td}|$ in single-top production at $p e^{\mp}$ colliders}, arXiv:2405.12400 [hep-ph].
\bibitem{Clerbauxjhep2019} B. Clerbaux, W. Fang, A. Giammanco, R. Goldouzian, \emph{Model-independent constraints on the CKM
matrix elements $|V_{tb}|$, $|V_{ts}|$ and $|V_{td}|$}, J. High Energ. Phys. 2019, 22 (2019). https://doi.org/10.1007/JHEP03(2019)022. 

\bibitem{Diaz:2002uk}
R.~A.~Diaz, R.~Martinez and J.~A.~Rodriguez,
Phys. Rev. D \textbf{67} (2003), 075011
doi:10.1103/PhysRevD.67.075011
[arXiv:hep-ph/0208117 [hep-ph]].


\bibitem{Aoyamag2} T. Aoyama et al., \emph{The anomalous magnetic moment of the muon in the Standard Model}, Phys. Rept. 887 (2020) 1
. https://doi.org/10.1016/j.physrep.2020.07.006.

\bibitem{Abimug2} B. Abi et al. [Muon g-2], \emph{Measurement of the Positive Muon Anomalous Magnetic Moment to 0.46 ppm}, Phys. Rev. Lett. 126 

\bibitem{Muong-2:2023cdq}
D.~P.~Aguillard et al., Muon g-2 Collab., 
Phys. Rev. Lett. \textbf{131}, 161802 (2023). 


\bibitem{FermilabLattice:2019ugu}
C.~T.~H.~Davies \textit{et al.} [Fermilab Lattice, LATTICE-HPQCD and MILC],
Phys. Rev. D \textbf{101} (2020) no.3, 034512
doi:10.1103/PhysRevD.101.034512
[arXiv:1902.04223 [hep-lat]].


\bibitem{Borsanyi:2020mff}
S.~Borsanyi, Z.~Fodor, J.~N.~Guenther, C.~Hoelbling, S.~D.~Katz, L.~Lellouch, T.~Lippert, K.~Miura, L.~Parato and K.~K.~Szabo, \textit{et al.}
Nature \textbf{593} (2021) no.7857, 51-55
doi:10.1038/s41586-021-03418-1
[arXiv:2002.12347 [hep-lat]].

\bibitem{Lehner:2020crt}
C.~Lehner and A.~S.~Meyer,
Phys. Rev. D \textbf{101} (2020), 074515
doi:10.1103/PhysRevD.101.074515
[arXiv:2003.04177 [hep-lat]].


\bibitem{Aubin:2019usy}
C.~Aubin, T.~Blum, C.~Tu, M.~Golterman, C.~Jung and S.~Peris,
Phys. Rev. D \textbf{101} (2020) no.1, 014503
doi:10.1103/PhysRevD.101.014503
[arXiv:1905.09307 [hep-lat]].


\bibitem{Wittig:2023pcl}
H.~Wittig,
[arXiv:2306.04165 [hep-ph]].








\bibitem{Crivellin:2015hha}
A.~Crivellin, J.~Heeck and P.~Stoffer, \emph{A perturbed lepton-specific two-Higgs-doublet model facing experimental hints for physics beyond the Standard Model},Phys. Rev. Lett. 116 (2016) no.8, 081801. https://doi.org/10.1103/PhysRevLett.116.081801.

\bibitem{Barr:1990vd}
S.~M.~Barr and A.~Zee, \emph{Electric Dipole Moment of the Electron and of the Neutron}, Phys. Rev. Lett. 65, 21-24 (1990) [erratum: Phys. Rev. Lett.65, 2920 (1990)]. https://doi.org/10.1103/PhysRevLett.65.21.

\bibitem{Crivellin:2013wna}
A.~Crivellin, A.~Kokulu and C.~Greub, \emph{Flavor-phenomenology of two-Higgs-doublet models with generic Yukawa structure}, Phys. Rev. D 87 (2013) no.9, 094031. https://doi.org/10.1103/PhysRevD.87.094031.
\end{thebibliography}
\end{document}